\documentclass[prb,twocolumn,aps,showpacs,10pt]{revtex4}
\usepackage{graphicx}
\usepackage{bm}
\usepackage{amssymb} 
\usepackage{color}

\begin{document}

\title{Decay of the rotary echoes for the spin of a nitrogen-vacancy center in diamond}

\author{V. V. Mkhitaryan and V. V. Dobrovitski}

\affiliation{Ames Laboratory, Iowa State University, Ames, Iowa
50011, USA}

\begin{abstract}
We study dynamics of the electron spin of a nitrogen-vacancy (NV)
center subjected to a strong driving field with periodically
reversed direction (train of rotary echoes). We use analytical and
numerical tools to analyze in detail the form and timescales of
decay of the rotary echo train, modeling the decohering spin
environment as a random magnetic field. We demonstrate that the
problem can be exactly mapped onto a model of spin 1 coupled to a
single bosonic mode with imaginary frequency. This mapping allows
comprehensive analytical investigation beyond the standard
Bloch-Redfield-type approaches. We explore the decay of the rotary
echo train under assumption of strong driving, and identify the
most important regimes of the decay. The analytical results are
compared with the direct numerical simulations to confirm
quantitative accuracy of our study. We present the results for
realistic environment of substitutional nitrogen atoms (P1
centers), and provide a simplified but accurate description for
decay of the rotary echo train of the NV center's spin. The
approach presented here can also be used to study decoherence and
longitudinal relaxation of other spin systems under conditions of
strong driving.
\end{abstract}
\pacs{76.30.Mi, 03.65.Yz, 76.30.-v, 76.60.Lz}

\maketitle

\section{Introduction}

Single electron spins in solids hold much promise as qubits for
quantum-based technologies, and as a platform for studying
fundamental problems of quantum mechanics. Among these systems,
the nitrogen-vacancy (NV) centers in diamond exhibit a set of
particularly desirable features: individual centers can be
initialized and read out optically, \cite{optread1,WJ,optread4,
RoblHans11} possess naturally long coherence times even at room
temperature, \cite{NV2, longcoh2, Maurer12} and can be controlled
\cite{nvmanip} using magnetic fields, \cite{magmanip1,
magmanip2, HansDobAwsch08, magmanip4}
optical excitations, \cite{optmanip1, optmanip2, Togan10,
Faraon12, optmanip5} and electric fields.
\cite{elmanip1,elmanip2,elsens} As a result, the NV centers have
attracted much attention as prospective qubits for quantum
information processing, \cite{Togan10,RoblHans11,
Maurer12,vdSar12,Bernien13,Dolde13,Faraon12,Child06,Capp09} and as
nanoscale sensors. \cite{magsens1, magsens2, magsens3, magsens4,
magsens5, nvsens1, nvsens2, nvsens3, nvnmr1, nvnmr2, elsens,
tempsens1, tempsens2, tempsens3}

Efficiency of the NV-based devices critically depends on the NV
spin coherence time, which is controlled by the coupling to the
spins of substitutional nitrogen atoms and/or to the bath of
$^{13}$C nuclear spins. In order to decouple the NV spin from the
decohering environment, many pulse-based dynamical decoupling
protocols have been investigated and proven efficient.
\cite{nvdd1, nvdd2, nvdd3, nvdd4, nvdd5, nvdd6, nvdd7, Capp09,
nvdd9} At the same time, an alternative approach based on the
continuous dynamical decoupling has been extensively investigated.
\cite{HansDobAwsch08, DFHA, HirAielCap12, Loretz13, Facchi04,
Bermudez11, Cai12, Xu13, Cai13, Belt13, London13, LaraouiMeriles,
AielHirCap12} Within this approach, strong resonant driving is
applied to the NV center. The resulting fast Rabi oscillations of
the NV spin, similarly to the flip-flops induced by the decoupling
pulses, average out the interaction with the environment, and
significantly extend the spin coherence time. But, under realistic
circumstances, the coherence time (i.e.\ the decay time of Rabi
oscillations) is often limited by the slow drift of the driving
field and/or small accidental detuning of the driving frequency
from exact resonance. Robustness of the continuous decoupling with
respect to these experimental imperfections can be greatly
improved by periodically inverting the driving field (switching
the driving field phase by 180$^\circ$). \cite{Solomon, Slichter,
HirAielCap12, LaraouiMeriles, AielHirCap12} The resulting signal
demonstrates a series of so-called rotary echoes, which decay much
slower than the regular Rabi oscillations, and may be utilized for
precise nanoscale sensing. \cite{HirAielCap12, LaraouiMeriles,
AielHirCap12}

Although the rotary echo protocol requires constant driving of the
NV spin, and therefore much larger dissipated power, it has an
additional advantage over the pulse-based dynamical decoupling: an
experimentalist can independently control both the strength of
driving and the driving reversal time. This freedom gives access
to many different dynamical regimes, so that the most advantageous
regime can be chosen for a given experiment.

The goal of this work is to investigate in detail, both
analytically and numerically, the form and the rate of the decay
of the rotary echoes, caused by the coupling of the NV center to
its spin environment. Influence of the relevant dilute dipolar-coupled
spin bath on the NV spin can be modelled as a random time-varying
magnetic field obeying the statistics of the Ornstein-Uhlenbeck
process (stationary Gaussian Markovian process).
\cite{HansDobAwsch08, DFHA, envou1, deLange, HirAielCap12,
Bermudez11} Previous studies of the rotary echo decay of the NV
spin, \cite{HirAielCap12, LaraouiMeriles, AielHirCap12} while
providing important insights, were restricted to low orders of the
cumulant expansion of the evolution operator, or to the case of
static bath.
Our study avoids these limitation. We demonstrate that the
original problem is equivalent to a model of a driven spin coupled
to a {\it single} imaginary-frequency bosonic mode, which can be
explored in depth using various techniques. We focus on the
experimentally interesting case of strong driving (which ensures
long-living rotary echoes), and analyze the most interesting
dynamical regimes, which could be useful for extracting the
parameters of the spin bath, and for extending the NV spin
coherence time for precise sensing.

We note that the analysis here can be applied to a wide variety of
other spins decohered by the dilute dipolar-coupled baths, such as
donors in silicon or magnetic ions in non-magnetic host crystals.
\cite{Pla, Morton, Morello, Bertaina} In particular, our
analytical approach can be useful for studying decoherence and
longitudinal relaxation of a spin subjected to a strong driving,
without resorting to standard Bloch-Redfield-type approximations
based on neglecting non-secular terms in the equations of motion.
This may be important for understanding complex environments,
which often lead to non-exponential decoherence and relaxation.

The paper is organized as follows. In the next Section  we
consider dynamics of a driven NV spin coupled to a dilute spin
bath, and formulate the corresponding central spin problem. In
Sec.~\ref{sec:decay}, we outline the derivation of the
Fokker-Planck equation for this problem, and its mapping on the
spin coupled to an imaginary-frequency bosonic mode.
In the same Section, we present the analysis of the evolution
operator in the case of strong driving, and discus different
dynamical regimes. Analytical results are compared with the
numerical simulations in Sec.~\ref{sec:numerics}. In
Sec.~\ref{sec:realNV} we extend our treatment to encompass the
realistic setup of the strongly driven NV spin, which is coupled
to several baths, and where the effect of the hyperfine coupling
is taken into account. Discussion of the results, and brief
analysis of the asymmetric protocol, are presented in
Sec.~\ref{sec:discussion}. Appendices provide technical details of
the analysis used in Sec.~\ref{sec:decay} and \ref{sec:realNV}.


\section{NV center coupled to the dilute spin bath}

\label{sec:description}

The NV center is a negatively charged defect, consisting of a
substitutional nitrogen atom and an adjacent vacant cite in
diamond. The orbital electronic ground state of this defect is
spin triplet $S=1$. The levels $m_S=0$ and $m_S=\pm 1$ are split
by $D=2.87$~GHz, and the spin  quantization axis (denoted below as
$z$) coincides with the [111] crystal axis. Static magnetic field
$B_0$ is often applied along $\hat{z}$ to lift the degeneracy of
the states $m_S=+1$ and $m_S=-1$. In order to control the NV
center's spin, Rabi driving is applied in resonance with the
transition between $m_{\text S}=0$ and $m_{\text S}=-1$, while
$B_0$ of order of few tens to few hundred Gauss suffices to detune
this transition far enough from the other resonance, between
$m_{\text S}=0$ and $m_{\text S}=+1$. Therefore an effective
description of the NV spin as a two-level system, $S_0=1/2$, is
adequate for typical experimental situations.

The NV electron spin is coupled to the nuclear spin of the NV's
own nitrogen atom ($I_0=1$ for $^{14}$N isotope, $I_0=1/2$ for
$^{15}$N isotope) via hyperfine interaction $A_0S^z_0I^z_0$, where
$A_0=-2\pi\cdot 2.16$~MHz for $^{14}$N and $A_0=2\pi\cdot
3.03$~MHz for $^{14}$N. \cite{Childress,Felton} The nuclear spin
relaxation is slow, so that for a single experimental run the
nuclear spin state is constant, but changes randomly between
different experimental runs.

The longitudinal (spin-lattice) relaxation of the NV spin is very
slow, the corresponding time $T_1$ is in the range of tens of
milliseconds at room temperature, and becomes much longer at lower
temperature \cite{WJ}. In contrast, the transverse dephasing time
$T^*_2$ is usually of the order of microseconds. For the systems
considered here, the transverse decoherence is caused mainly by
the spin bath, made of a large number of the surrounding
electronic spins of the substitutional N atoms (also known as P1
centers)\cite{HansDobAwsch08,Loubser}. Characteristic
dipole-dipole coupling of the spin of the N atom to the NV
center's spin is of the order of MHz, and the coupling between
different N spins is of the same order of magnitude. The
flip-flops between the N spin and the NV center's spin are
suppressed due to the large energy mismatch \cite{HansDobAwsch08}.
However, the mutual flip-flops between the N spins are generally
allowed, and lead to fluctuations of the dipolar field created by
the spin bath on the NV center. Since the long-range dipolar
interaction couples each N spin to hundreds of other N spins and
only to one NV center, and all the couplings are comparable,
dynamics of the flip-flops is barely affected by the state of the
NV center (negligible back-action)
\cite{HansDobAwsch08,DFHA,nvdd4,nvdd1,Bermudez11}. Such a bath can
be described in a mean-field manner, as a random time-varying
magnetic field $B(t)$ acting on the NV spin, in a spirit of
earlier works on magnetic resonance \cite{Anderson1, Anderson2,
Kubo1, KlauderAnderson}. Such a description agrees very well with
both direct numerical simulations and with experimental results.
\cite{HansDobAwsch08,nvdd1,DFHA,DFAHPRB,deLange}

Note that the random-field approximation is justified when the
gyromagnetic ratios of the spins are not drastically different,
e.g.\ for the nuclear spin decohered by the bath of nuclear spins,
or electronic spin decohered by the bath of electronic spins. When
the electronic spin is decohered by the bath of nuclear spins, the
back-action from the central spin to many bath spins may become
strong compared to the coupling within the bath. In this case
decoherence of the central spin may be governed by a mechanism
similar to electronic spin echo envelope modulation (ESEEM)
\cite{Slichter}, modified by the many-spin nature of the bath, and
other methods must be used to analyze the decoherence dynamics
\cite{Witzel05, RBLiu07, Saikin07, Cywinski09, Maze08b, CoishLoss}

As a result, the influence of the N atom bath on the NV spin is
described with the simplified dephasing Hamiltonian $B(t)S^z_0$,
where the random field $B(t)$ is assumed to be Gaussian (because
many nitrogen atoms contribute to the field $B(t)$), stationary
(since the back-action on the bath is negligible), and Markovian
(due to the large size of the bath and negligible back-action)
stochastic process. Such a noise field is represented by an
Ornstein-Uhlenbeck (O-U) random process \cite{Kampen}, with the
correlation function
\begin{equation}
\label{OUcorr}
\langle B(0)B(t)\rangle=b^2 \exp{(-R|t|)},
\end{equation}
where $b$ is the rms of the fluctuating random field, and $R$ is
the correlation decay rate. In fact, since the P1 centers can have
different orientations and different internal states
\cite{HansDobAwsch08,deLange}, the actual random field created by
the N atom bath is a sum of six O-U processes, which are
independent with a good accuracy, each having its own parameters
$b$ and $R$. This will be taken into account in
Sec.~\ref{sec:realNV}.

Since the carrier frequency of the Rabi driving field is in the
GHz range, the effect of the driving field can be described in the
rotating frame, by neglecting the counter-rotating terms
\cite{Slichter}. Restricting our consideration to the two relevant
levels $m_S=-1$ and $m_S=0$ (with the third level $m_S=+1$ staying
idle), we arrive at the Hamiltonian describing the system under
consideration
\begin{equation}
\label{Ham2}
H= B_1(t) S^z + h(t) S^x,
\end{equation}
where $S_{z}=|m_S=0\rangle\langle m_S=0|-|m_S=-1\rangle\langle
m_S=-1|$, $S_x=|m_S=0\rangle\langle m_S=-1|+|m_S=-1\rangle\langle
m_S=0|$ are the effective spin operators in the relevant two-level
subspace, $h(t)$ is the Rabi driving whose magnitude $h$ is large
($h\gg b,\ R$, etc.) and whose sign is periodically inverted, and
$B(t)$ is the total detuning from the driving frequency: it
includes the possible static detuning, the quasi-static hyperfine
field, and the dynamic O-U field $B(t)$ created by the spin bath.

\section{Decay of the rotary echoes under strong driving}

In order to analyze the long-time dynamics, and to correctly
account for the long-time accumulation effects, we first consider
the case where the static detuning is zero (driving is in exact
resonance with the transition $m_S=0\leftrightarrow m_S=-1$), and
the on-site hyperfine coupling is omitted. Moreover, in this
Section, we restrict our consideration to a single O-U random
field; the effect of the realistic bath comprising six different
O-U processes is considered in Sec.~\ref{sec:realNV}.
Correspondingly, we consider the Hamiltonian
\begin{equation}
\label{eq:hamstart}
H= B(t) S^z + h(t) S^x,
\end{equation}
where $h(t)$ is the driving, whose direction is periodically
switched between $+x$ and $-x$, and $B(t)$ is the O-U random
process.

Our goal is to study the long-time dynamics of the central spin,
and in spite of the large magnitude of $h$, the impact of the
field $B(t)$ can accumulate over long periods of time and over
many reversals of the driving field $h(t)$. Taking into account
this accumulation effect is not straightforward. For instance, let
us consider the interval between two reversals, when the driving
$h$ is constant. A standard way of treating this case is to
perform a unitary rotation, which turns the Hamiltonian
(\ref{eq:hamstart}) to the form
\begin{equation}
H \approx [h + B^2(t)/(2 h)] {\bar S}^x
\end{equation}
which is valid up to $1/h^2$, and where ${\bar S}^x$ differs from
$S^x$ by the terms of order of $1/h$. The difference between
${\bar S}^x$ and $S^x$ is usually neglected, since it remains
small at long times, while the phase fluctuations caused by the
factor $B^2(t)/(2 h)$ accumulate. However, when the direction of
the driving is constantly reversed, the difference between ${\bar
S}^x$ and $S^x$ also accumulates after every reversal, and
eventually becomes important. In addition, the fluctuating field
$B(t)$ induces incoherent transitions between the eigenstates of
$S^x$, which are usually treated within the Bloch-Redfield theory
\cite{Slichter} and its generalizations \cite{Kosloff,CohTan}, and
which lead to exponential relaxation with the rate of order of
$1/h^2$. The process of such relaxation in the case when the field
$h$ is periodically reversed has not, to our knowledge, been
investigated before, but the formalism presented below directly
accounts for such processes.

\label{sec:decay}

\subsection{Fokker-Planck equations and mapping onto the spin-$1$
model}


To study the time evolution of the central spin, governed by the
Hamiltonian Eq. (\ref{Ham2}), we write the spin's density matrix
as $\varrho(t)=\frac12[1+ m_x\sigma^x +m_y\sigma^y +m_z\sigma^z]$,
where $\sigma$'s are the Pauli matrices, and $|{\bf m}(t)|\leq 1$.
Its evolution is given by $\dot{\varrho}=i[\varrho, H]$, entailing
the stochastic differential equation,
\begin{equation}
\label{stdiffeq} \frac {d m_\mu}{dt}=F_\mu\bigl({\bf m},
B(t)\bigr),\quad \mu=x,\,y,\,z,
\end{equation}
where
\begin{equation}
\label{Fs} F_x=-B(t)m_y,\,\, F_y=B(t)m_x-h\,m_z,\,\, F_z= h\,m_y.
\end{equation}
Dynamics of the average values of the variables $m_\mu(t)$ can be
analyzed using the method suggested by R.\ Kubo
\cite{Kubo,Kampen}.

Since $B(t)$ is Markovian, so is the joint process described by the
variables $({\bf m}, B)$. Then its joint probability density,
${\cal P}({\bf m}, B,t)$, obeys the following
stochastic Liouville equation \cite{Kubo}:
\begin{equation}
\label{SLE} \frac{\partial {\cal P}({\bf m}, B,t)}{\partial t}=
-\sum_\mu\frac{\partial\bigl[F_\mu{\cal P}\bigr]}{\partial m_\mu}
+ R\partial_B\bigl[B{\cal P}\bigr]+ Rb^2\partial^2_B{\cal P}.
\end{equation}
Since $\bf F$ is linear in $\bf m$, one can directly obtain the
equations of motion for the marginal averages $v_\mu(B,t)= \int
m_\mu {\cal P}({\bf m}, B,t)d{\bf m}$. Multiplying Eq.~(\ref{SLE}) by
$m_\mu$ and performing integration, we obtain the system of
coupled equations
\begin{eqnarray}
\label{F-P} &&\partial_tv_x(B,t)=-Bv_y+ R\partial_B[Bv_x]+
Rb^2\partial^2_Bv_x, \nonumber\\
&&\partial_tv_y(B,t)=Bv_x-hv_z+ R\partial_B[Bv_y]+
Rb^2\partial^2_Bv_y,\qquad\nonumber\\
&&\partial_tv_z(B,t)=hv_y+ R\partial_B[Bv_z]+ Rb^2\partial^2_Bv_z.
\end{eqnarray}
These equations should be solved with the initial conditions
$v_\mu(B,0)= P_0(B)m_\mu(0)$. Then the average values $\langle
m_\mu(t)\rangle$ can be obtained by straightforward integration,
since $\langle m_\mu(t)\rangle=\int v_\mu(B,t)dB$.

The system (\ref{F-P}) is complex, and we are not aware of any
means of obtaining an exact analytical solution. However, we can
re-formulate the problem in terms of a quantum spin 1 coupled to
an oscillator with imaginary frequency. After such a
transformation, we will be able to use the tools of standard
quantum mechanics to explore the regime of strong driving, and
derive the approximation valid at arbitrarily long times.

For convenience of notation, let us introduce the $so(3)$
generators
\begin{equation}
\label{O3} \hat{g}_x= {\small \left(\!\begin{array}{ccc}
0&0&0\\
0&0&-1\\
0&1&0
\end{array}\!\right)},\quad \hat{g}_y=
{\small \left(\!\begin{array}{ccc}
0&0&1\\
0&0&0\\
-1&0&0
\end{array}\!\right)},\quad
\hat{g}_z\!=\! {\small \left(\!\begin{array}{ccc}
0&-1&0\\
1&0&0\\
0&0&0
\end{array}\!\right)\! ,}
\end{equation}
and dimensionless variables
\begin{equation}
\label{dimless} \tilde{t}=\sqrt{2}b\,t,\quad \xi=\frac
B{\sqrt{2}b},\quad \omega=\frac h{\sqrt{2}b}, \quad\rho=\frac
R{\sqrt{2}b}.
\end{equation}
A crucial step towards the solution of our problem is the
observation that, upon introducing $\hat{\psi} (\xi,\tilde{t})=
e^{\xi^2/2}{\bf v}(\xi,\tilde{t})$, Eq. (\ref{F-P}) acquires the
form
\begin{equation}
\label{dyn}
-\partial_{\tilde{t}}\,\hat{\psi}=\hat{H}\hat{\psi},\qquad
\hat{H}=\rho\, a^\dag a - \omega\hat{g}_x- \frac{a+
a^\dag}{\sqrt{2}}\hat{g}_z,
\end{equation}
where $a^\dag=(-\partial_\xi+\xi)/\sqrt{2}$ and
$a=(\partial_\xi+\xi)/\sqrt{2}$ form a standard pair of the
creation/annihilation operators of a harmonic oscillator. The
dynamical equation (\ref{dyn}) is equivalent to a Schr\"odinger
equation for a spin coupled to an imaginary-frequency oscillator:
\begin{equation}
i\partial_{\tilde{t}}\,\hat{\psi}=\hat{G}\hat{\psi},\qquad
\hat G = -i\rho\, a^\dag a + \omega\hat{s}_x + \frac{a+
a^\dag}{\sqrt{2}}\hat{s}_z,
\end{equation}
where $s_{\mu}=ig_\mu$ can be viewed as spin operators of a spin
$s=1$. The above mentioned initial condition translates into
$\hat{\psi}(\xi,0)= {\bf m}(0)\otimes \exp(-\xi^2/2)$, while the
averages become $\langle m_\mu(\tilde{t})\rangle=\int
\frac{d\xi}{\sqrt{\pi}} e^{-\xi^2/2}\psi_\mu(\xi,\tilde{t})$. Thus
the problem of finding the $z$-component of the central spin,
which is initially directed along the $z$-axis, reduces to the
evaluation of the matrix element
\begin{equation}
\label{viamatel} \langle S^z(t)\rangle=\frac12\langle0_z|
\exp[-\tilde{t}\hat{H}]|0_z\rangle = \frac12\langle0_z|
\exp[-i\tilde{t}\hat{G}]|0_z\rangle,
\end{equation}
where $\langle0_z|=(0,0,1) \otimes\langle0|$, and
$\langle0|=\pi^{-1/4}\exp(-\xi^2/2)$ stands for the ground state
of the oscillator mode.

Thus, we {\it exactly\/} mapped the stochastic model (\ref{Ham2})
onto the quantum mechanical one, given by Eq.~\ref{dyn}. The
operators $\hat{s}_\mu$ satisfy the commutation relations of the
$su(2)$  algebra,
$[\hat{s}_\mu,\hat{s}_\nu]=i\epsilon_{\mu\nu\lambda}
\hat{s}_\lambda$, and the Casimir operator is $\sum
\hat{s}_\mu^2=2$. Therefore, Eq.~\ref{dyn} is the Schr\"{o}dinger
equation for a spin $s=1$, subjected to a strong magnetic field
$\omega$ along the $x$-direction, and linearly coupled to a
harmonic oscillator with an {\it imaginary} frequency $-i\rho$.

Although this model, as far as we know, is not exactly solvable,
but the regime of interest is characterized by a large parameter
$\omega\gg 1$. This can be used to construct the perturbative
series which would be correct at arbitrarily long times. Taking
this advantage, we find the temporal evolution of rotary echo
amplitude in the next subsection.

\subsection{Description of the rotary echoes}

Rotary echoes are produced by periodic reversals of the driving
field. The simplest protocol is to switch the driving phase so
that $h(t)=h$ for $0<t<2T$, then $h(t)=-h$ for $2T<t<4T$, then
back to $h(t)=h$ for $4T<t<6T$, etc. In this protocol, the unit
($h(t)=h$ for $0<t<2T$ and $h(t)=-h$ for $2T<t<4T$) is repeated
with the period $4T$. However, this protocol does not produce good
protection against decoherence, as explained at the end of
Sec.~\ref{sec:discussion}. Therefore, we omit the detailed
analysis of this protocol here.

Instead, we focus on the improved version of the rotary echo
protocol, which is also periodic, but where the basic unit is
symmetrized: $h(t)=h$ for $0<t<T$, then $h(t)=-h$ for $T<t<3T$,
and again $h(t)=h$ for $3T<t<4T$. Such a unit is repeated with the
period $4T$, and provides good protection against the random noise
field $B(t)$.

This symmetric $N$-cycle driving with the
reversal period $4\tau$ is described by the evolution
operator $\hat{U}(N)=U^N(\tau|2\tau|\tau)$, where
\begin{equation}
\label{Usym}
U(\tau|2\tau|\tau)=e^{-\tau \hat{H}_{+}} e^{-2\tau
\hat{H}_{-}}e^{-\tau \hat{H}_{+}}
\end{equation}
is the evolution operator for a single cycle. Here
$\tau=\sqrt{2}bT$ is the dimensionless quarter-period, while the
Hamiltonians $\hat{H}_{\pm}=\hat{H}(\pm\omega)$ correspond to
opposite signs of the driving field amplitude. The rationale
behind the effectiveness of this protocol is the cancellation of
phases accumulated by the central spin for the time $T$ before and
for the time $T$ after the reversal.
If $T$ is short enough, this cancellation is almost
complete, and the dephasing is strongly reduced.

The amplitude of the $N$-th rotary echo is given by the average
\begin{equation}
\label{SviaME} \langle S^z(N)\rangle=
\frac12\langle0_z|U^N(\tau|2\tau|\tau)|0_z\rangle.
\end{equation}
We calculate $\langle S^z(N)\rangle$ utilizing the large value of
$\omega$, but keeping in mind that $N$ can be as large as
$\omega$, i.e. we assume that  $N\sim \omega$, although
$N/\omega^2\ll 1$.

More formally, let us denote the eigenvectors of
$U(\tau|2\tau|\tau)$ by $|u_i\rangle$ and the corresponding
eigenvalues by $e^{-\lambda_i}$, so that $U^N=\sum_i
e^{-\lambda_iN}|u_i\rangle\langle u_i|$. If we find the
eigenvectors and eigenvalues to some accuracy, then the error in
the eigenvalue accumulates with $N$, while the error in the
eigenvector does not accumulate (taking into account that
$\text{Re} \lambda_i\ge 0$). Hence we are going to approximate
$|u_i\rangle$ with $|\bar{u}_i\rangle$, and $\lambda_i$ with
$\bar{\lambda}_i$ such that $(|u_i\rangle- |\bar{u}_i\rangle)\sim
1/\omega$ and $(\lambda_i-\bar{\lambda}_i) \sim1/\omega^2$. Then
the operator, $\bar{U}^N=\sum_i
e^{-\bar{\lambda}_iN}|\bar{u}_i\rangle \langle \bar{u}_i|$ will
provide the desired approximation to evaluate $\langle
S^z(N)\rangle$.

A standard approach is to use perturbative treatment of the model
Eq. (\ref{dyn}) over small $1/\omega$. However this approach is
not justified here, because the parameter $\rho$ can be either
smaller or larger than $1/\omega$, without restriction. Forgetting
this fact would lead to suppressed denominators in the asymptotic
expansion over $1/\omega$. Instead, we approximately transform the
pseudo-Hamiltonian $\hat{H}$, given by Eq.~(\ref{dyn}), to a more
convenient form, by applying a sequence of small
Schrieffer-Wolff-like transformations. In Appendix A we show that,
within the necessary accuracy, the time evolution operator
(\ref{Usym}) is given by
\begin{equation}
\label{approxU} U\approx W_0^{\dag}\left[e^{-\tau
\hat{h}_{+}}W_0^2e^{-2\tau \hat{h}_{-}}W_0^{\dag\,2} e^{-\tau
\hat{h}_{+}}\right]W_0,
\end{equation}
where we have the small rotation
\begin{equation}
\label{W0W1} W_0=\exp\!\left[\frac{a+a^\dag} {\sqrt{2}\,\omega}
\hat{g}_y \right],
\end{equation}
and $\hat{h}_{\pm}=\hat{h}(\pm\omega)$ with the reduced
pseudo-Hamiltonian,
\begin{equation}
\label{hsmall} \hat{h}(\omega)= \rho\, a^\dag a - \omega\hat{g}_x
- \frac{(a+ a^\dag)^2}{4\omega} \hat{g}_x.
\end{equation}
Equation (\ref{approxU}) can be understood as follows. The
pseudo-Hamiltonian $\hat{H}$, Eq. (\ref{dyn}), defines the motion
of $\bf \hat{s}$ in an effective magnetic field almost parallel to
the $x$ axis, but slightly tilted to the $z$ direction (which
tilting is conditioned on the state of the oscillator). To the
first order, this tilting can be taken into account by means of a
small rotation in the $x$--$z$ plane, which adjusts the spin's
effective quantization axis towards the direction of the effective
magnetic field. Then the pseudo-Hamiltonian $\hat{H}\equiv
\hat{H}_+$ reduces to $\hat{h}_+$, whereas $W_0$ is the operator
performing the necessary small rotation. When the direction of
driving switches to $-h$, the direction of the effective field
changes to the almost (but not exactly!)  opposite, and the
quantization axis assumes new direction; the corresponding
transformation is implemented by $W_0^\dag$, and the resulting
Hamiltonian is $h_-$. The arrangement of operators $W_0$,
$W_0^\dag$, and $\hat{h}_{\pm}$ in Eq. (\ref{approxU}) is a
consequence of the fact that while $\hat{H}_+$ goes to $\hat{h}_+$
via the $W_0$-rotation, $\hat{H}_-$ transforms to $\hat{h}_-$ by
means of the $W_0^\dag$-rotation. The detailed derivation of
Eq.~\ref{approxU} and analysis of its accuracy are given in
Appendix A.

It is easy to see that operators $W_0^{\dag}$ and $W_0$ outside
the rectangular brackets of Eq. (\ref{approxU}) transform the
eigenvectors of $U$ by $\sim1/\omega$, but do not affect its
eigenvalues. Their contribution \cite{footnote} to the matrix
element Eq. (\ref{SviaME}) does not accumulate with $N$, so that
we can neglect these operators. In contrast, the operators $W_0^2$
and $W_0^{\dag\,2}$ inside the rectangular brackets of Eq.
(\ref{approxU}) can not be simply neglected, as they affect both
eigenvectors and eigenvalues of $U$, and accumulate with
increasing $N$. In Appendix C we carefully analyze these terms and
show that they, in essence, contribute $\sim1/\omega^2$ to the
eigenvalues of $U$. Therefore, they can be neglected provided that
$N/\omega^2\ll 1$. Our approach allows to go beyond that
restriction, and gain insights into the case $N/\omega^2\sim 1$,
but we postpone this until Sec.~\ref{sec:largerN}.

Therefore, for the purposes of this Section, the time evolution
operator (\ref{Usym}) is well approximated with $U_0=e^{-\tau
\hat{h}_{+}}e^{-2\tau \hat{h}_{-}} e^{-\tau \hat{h}_{+}}$, and the
decay of the rotary echoes is described by
\begin{equation}
\label{SWres} \langle
S^z(N)\rangle\approx\frac12\langle0_z|U_0^N\!|0_z\rangle.
\end{equation}
The matrix element (\ref{SWres}) is evaluated by solving the
eigenvalue problem for $U_0$ and expanding $|0_z\rangle$ over the
complete set of the eigenvectors of $U_0$, see Appendix B. To
outline the result, we introduce the real positive quantity $\psi$
defined as
\begin{equation}
\label{psi} \cosh\psi=\cosh\! 2\tau P\cosh\!2\tau
P^*+\frac{\rho^2}{|P|^2}\sinh\! 2\tau P\sinh\!2\tau P^*\!,
\end{equation}
where $P=\sqrt{\rho^2-i\rho/\omega}$, and the complex-valued
quantity $q_z$ defined as
\begin{eqnarray}
\label{Qbig} q_z=\text{Re}\left[(2\rho\omega- i)\frac{\sinh2\tau
P\cosh2\tau P^*}{\omega
P\sinh\psi}\right]&&\nonumber\\
- \frac{\rho\sinh2\tau P^*(\cosh2\tau P-1)}{2\omega^2
P|P|^2\sinh\psi},&&
\end{eqnarray}
where stars mean complex conjugation.
In terms of these quantities, our result becomes:
\begin{equation}
\label{resulttext} \langle
S^z(N)\rangle\simeq\frac12\text{Re}\frac{e^{2N\rho\tau}}
{\sqrt{\cosh N\psi+ q_z\sinh N\psi}},
\end{equation}
where the restriction $N/\omega^2\ll1$ is presumed. This analytic
form sets two different  dynamical regimes: the short-$\tau$
regime, where $|P|\tau\ll1$, and the long-$\tau$ regime, where
$|P|\tau\gg1$. Due to the presence of hyperbolic functions in the
definitions of $\psi$ and $q_z$, the crossover between the two
regimes occurs quickly. In the next subsection, we study the
behavior of the solution (\ref{resulttext}), utilizing asymptotic
forms of $\psi$ and $q_z$.

\subsection{Different regimes of the rotary echo decay}

\begin{figure}
\centerline{\includegraphics[width=95mm,angle=0,clip]{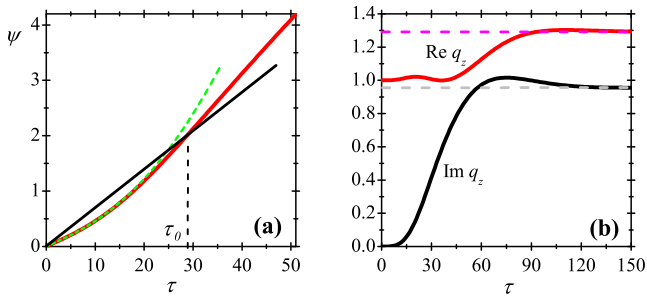}}
\caption{(Color online) (a): The quantity $\psi$ is plotted versus
$\tau$ with red (dark gray) from Eq. (\ref{psi}) for a slow bath,
$\rho=0.01$ and $\omega=10$. Its short-$\tau$ asymptote is plotted
with green (gray) dashed line from Eqs. (\ref{psiasympsmall}). The
line through the origin intersects $\psi$ at $\tau=\tau_0$. (b):
The real and imaginary parts of $q_z$ are plotted versus $\tau$
from Eq.~(\ref{Qbig}) (continuous lines), for the same bath. The
large-$\tau$ asymptotes, plotted from Eq. (\ref{Qasymplg}), are
shown with the dashed lines.} \label{slowbathpic}
\end{figure}

Despite its cumbersome form, the behavior of the quantity $\psi$
is not very complex. It can be approximated well by its asymptotic
forms,
\begin{eqnarray}
\label{psiasympsmall} &&\psi=4(\rho\tau)\sqrt{1+\tau^2/3\omega^2},
\qquad\qquad \tau<|P|^{-1},\\
&&\psi=4\tau\text{Re}P +\ln\frac12\bigl(1+
\rho^2/|P|^2\bigr),\quad \tau>|P|^{-1},\qquad
\label{psiasymplarge}
\end{eqnarray}
with a rather narrow crossover region between them. For $q_z$ we
have
\begin{eqnarray}
\label{Qasymp} &&q_z=\frac{1+\frac16\left(\tau/\omega\right)^2 -
\frac i6(\rho\tau)\left(\tau/\omega\right)^3}
{\sqrt{1+\frac13\left(\tau/\omega\right)^2}},
\qquad \tau<|P|^{-1},\qquad\\
&&q_z=\frac{2\rho\omega-i}{2\omega P},\quad\qquad\qquad\qquad
\tau>|P|^{-1}.\qquad \label{Qasymplg}
\end{eqnarray}
These asymptotes, as well as Eqs. (\ref{psi}) and (\ref{Qbig}),
predict qualitatively different behavior of $\psi$ and $q_z$ for
fast baths, with $\rho\gg1/\omega$, and for slow baths, where
$\rho\ll1/\omega$. For the fast baths, the above quantities
deviate from their asymptotic values, $\psi_0=4\tau\text{Re}\{P\}$
and $q_{z,0}=1$, only to the order, $(\rho\omega)^{-2}$, or even
less. For slow baths, their behavior is more complex, see Fig.
\ref{slowbathpic}.

At large times $\tilde t=4N\tau\gg 1$, Eq.~\ref{resulttext}
predicts exponential decay $[2(1+q_z)]^{-\frac12}\exp(-\Gamma
\tilde t)$, where the rate $\Gamma=\psi/8\tau-\rho/2$ depends on
$\tau$. To estimate the efficiency of the symmetric rotary-echo
protocol, we fix the total interrogation time $\tilde t$ and look
at the dependence of the decay on $\tau$. With $\tau$ larger than
$|P|^{-1}$, both $\psi$ and $q_z$ reach their asymptotic forms
very quickly. Once $\psi$ becomes linear in $\tau$, and $q_z$
becomes nearly constant, the decay becomes almost insensitive to
$\tau$ or $N$. The only dependence on $\tau$ that still remains in
$\Gamma$, is due to the negative logarithm in Eq.
(\ref{psiasymplarge}). Its contribution to $\Gamma$ changes from
$-\ln2/8\tau$ (for slow bath), to zero (for fast bath). Even for
the slow bath, however, this is inessential compared to the
background of larger $\Gamma\approx\frac12\text{Re}P$.

The overall suppression of the decay rate, offered by the
symmetric rotary echo in comparison with the ordinary Rabi driving
\cite{DFHA} is exactly this contribution. This means that the
cancellation of the accumulated phases expected from the rotary
echo protocol, is incomplete for $\tau>|P|^{-1}$ because of the
evolving bath. Therefore, this regime is inefficient, and for
better protection of the echo, one needs to switch the driving
more frequently, ensuring $\tau<|P|^{-1}$. This is clearly visible
in Fig. \ref{slowbathpic} a. For a particular $\tau_0$, the decay
rate is given by the slope of the line through the origin,
intersecting $\psi$ at $\tau=\tau_0$. Obviously, smaller $\Gamma$
is reached with shorter $\tau_0$.

The symmetric rotary echo protocol is highly efficient when the
reversal period remains within the domain of small $\tau$,  for all kind
of baths. For fast bath, $\rho\gg1/\omega$, the decay is
exponential with a strongly suppressed rate,
\begin{equation}
\label{smallesttau} \langle S^z(N)\rangle \simeq\frac12 \exp
\left[-\frac{\rho\tau^3}{3\omega^2}N \right]
\end{equation}
over the whole domain of small $\tau$. The above result is derived
by setting $q_z=1$, which follows from Eq. (\ref{Qasymp}), using
the fact that, in this domain, $\rho\gg1/\omega$ ensures
$\tau\ll\omega$.

For moderate to slow baths $\rho\lesssim1/\omega$, however, the
domain $\tau<|P|^{-1}$ is further divided into two regions. When
$\tau$ is very small, $\tau\ll\omega$, the rotary echo maxima are
given by Eq. (\ref{smallesttau}). For larger $\tau$, when
$\omega\ll\tau<|P|^{-1}$, the decay is influenced by the growing
$|q_z|$. Interestingly, the echo decay in this region can become
non-exponential, given by
\begin{equation}
\label{nonexpdec} \langle S^z(N)\rangle \simeq\frac12
\left[1+\frac23(\rho\tau)\left(\frac\tau\omega\right)^2 N
\right]^{-\frac12}.
\end{equation}
This non-exponential decay takes place until $N<1/\psi$. For larger
values of $N$, the decay becomes exponential again,
\begin{equation}
\label{expdecreent} \langle S^z(N)\rangle\simeq
\frac{\exp\left(-\left[ \frac1{\sqrt{3}}
\left(\frac\tau\omega\right)
 -1\right]4\rho\tau N\right)}
{\sqrt{2+\frac1{\sqrt{3}}\left(\frac\tau\omega\right)}}.
\end{equation}
Note, however, that in this region it is possible that
$(\rho\tau)\bigl(\tau/\omega\bigr)^2>1$, so that the echo
amplitude decays almost completely already before reentering the
exponential regime described by Eq.~\ref{expdecreent}.

Eqs.~\ref{smallesttau}--\ref{expdecreent} summarize the behavior
of rotary echoes for all baths, subject to the restriction on the
number of the protocol periods $N/\omega^2\ll1$. In most
experimental situations it is realistic to tune the driving field
and set $\tau\ll\omega$, thus producing very long-lived rotary
echoes, with very slow decay given by Eq.~\ref{smallesttau}. On
the other hand, the non-exponential behavior can help gain more
insight into characteristics of the spin bath.

\subsection{Larger $N$}
\label{sec:largerN}

The result (\ref{resulttext}), and its short-$\tau$ forms,
Eqs.~\ref{smallesttau}--\ref{expdecreent}, are applicable when the
number of the driving reversals is not too large,
$N/\omega^2\ll1$. However, one may also be interested in the
rotary echoes at longer times, when $N/\omega^2\sim1$. As follows
from Eq.~\ref{smallesttau}, even at such long times one can still
observe well-defined rotary echoes, provided that $\rho\tau^3$ is
not too large, i.e.\ $\rho\tau^3\sim1$. In order to describe this
regime, we should find the corrections of order of $1/\omega^2$ to
Eq.~\ref{SWres}, which appear due to the factors $W_0^2$ and
$W_0^{\dag\,2}$, i.e.\ the corrections caused by the fact that the
quantization axis of the central spin changes its direction every
time the driving is reversed.

In Appendix D we show that for $\rho\tau\ll1$, this correction is
suppressed by the small factor $\rho\tau$, and can therefore be
neglected. Correspondingly, we restrict our consideration to the
less trivial case, when $\rho\tau\sim 1$. As before, we focus on
the experimentally relevant short-$\tau$ regime, $\tau<|P|^{-1}$.
These two conditions imply that the non-trivial corrections at
very large $N$ are relevant for fast baths where
$\rho\gg1/\omega$.

The $1/\omega^2$ corrections to Eq.~\ref{SWres} are calculated in
Appendix D. The result,
\begin{equation}
\label{CorrText} \frac{\Delta\langle S^z(N)\rangle}
{\langle0_z|U_0^N |0_z\rangle }=
-\frac{2N}{\omega^2}\tanh(\tau\rho)
\cos^2\bigl([\omega+1/4\omega]\tau\bigr),
\end{equation}
implies that the correction is negative (i.e.\ leads to faster
decay) and oscillates as a function of $\tau$. It also suggests
that the best decoupling is reached when $\tau$ is chosen amongst
the values,
\begin{equation}
\label{besttau} \tau_m=\frac{(2m+1)\pi} {2\omega+1/2\omega}
\end{equation}
with integer $m$, since for these values of $\tau$ the correction
disappears.

Besides, Eq.~\ref{resulttext} is obtained in Appendix B by
omitting the last term of $\hat{h}_{\pm}$, cf. Eq.~\ref{smSW},
which describes the longitudinal relaxation of the central spin
along the $x$-axis. Account of this term would multiply
Eq.~\ref{resulttext} by the exponential factor
$\exp(-N\rho\tau/\omega^2)$. Neglecting this term is legitimate
for $N/\omega^2\ll1$, but for larger $N$ we have to restore it.
Altogether, we arrive at the more accurate formula,
\begin{eqnarray}
\label{moreacur} \langle S^z(N)\rangle\simeq&&\!\!\!
\frac12\text{Re}\frac{\exp\bigl(
N\rho\tau[2-1/\omega^2]\bigr)}{\sqrt{\cosh N\psi+ q_z\sinh N\psi}}
\\
&&\times\left[1-\frac{2N}{\omega^2}\tanh(\tau\rho)
\cos^2\bigl([\omega+1/4\omega]\tau\bigr)\right]\!,\nonumber
\end{eqnarray}
which holds for $2N/\omega^2<1$, $\tau<|P|^{-1}$, and for
all kinds of baths.

\section{Numerical simulations}

\label{sec:numerics}

\begin{figure}
\vspace{-0.5cm}
\centerline{\includegraphics[width=90mm,angle=0,clip]{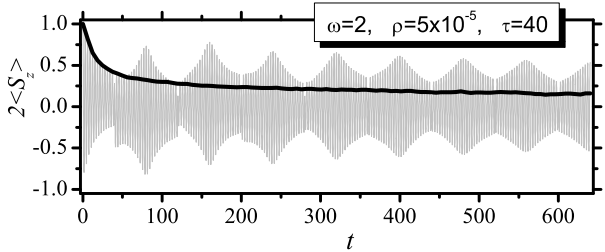}}
\vspace{-0.5cm} \caption{(Color online) Simulations results for
the central spin oscillations in a quasi-static bath,
$\rho=5\times 10^{-5}$. Individual oscillations caused by the
periodically reversed driving are shown in gray, as a function of the total time
$\tilde t=4N\tau$. The envelope of the conventional Rabi
oscillations for the same bath is plotted with black. Rotary
echoes are well pronounced and decay much slower than the Rabi
oscillations.} \label{pronounced}
\end{figure}

To gain better quantitative insights into the different regimes of
the rotary echo decay, and to check our analytical results, we
performed direct numerical simulation of the central spin
subjected to the rotary echo driving and the O-U magnetic noise.
In all cases we found excellent quantitative agreement between the
analytical results described above and the simulations.

\begin{figure}
\vspace{-0.5cm}
\centerline{\includegraphics[width=95mm,angle=0,clip]{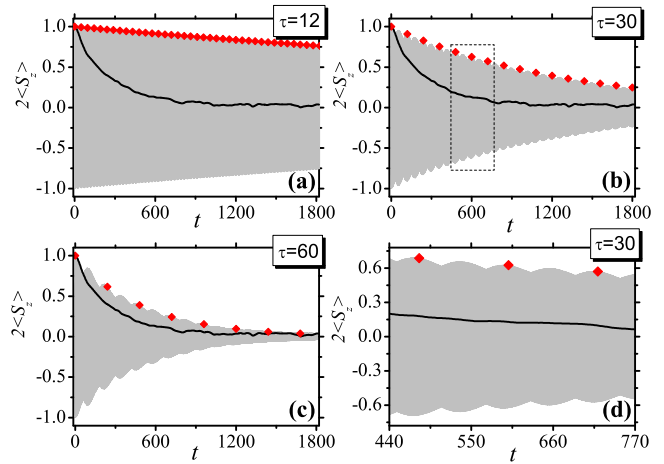}}
\vspace{-0.5cm} \caption{(Color online) Numerical simulations for
the slow bath, $\omega=20$ and $\rho=0.005$. Horizontal axes
represent the total time $\tilde t=4N\tau$. Individual
oscillations within the rotary echo signal (gray) have very high
frequency, and are not well resolved in the picture. Analytical
predictions for the rotary echo amplitudes (red diamonds), as
obtained from Eq.~\ref{resulttext}, perfectly match the
simulations. The envelope of the conventional Rabi oscillations
with the same parameters is shown with the black solid line.
Panels (a)--(c) correspond to three different values of $\tau$:
(a) $\tau=12$; (b) $\tau=30$, and (c) $\tau=60$. The panel (d) is
a zoom-in of the region marked as a dotted rectangle from the
panel (b). Note that the rotary echo protocol is very efficient
for short $\tau$, but its efficiently drops, and becomes
comparable to a standard Rabi oscillation, as $\tau$ approaches
$|P|^{-1} \approx 63.3$.} \label{slowgraphs}
\end{figure}

Rotary echoes are particularly well pronounced for quasi-static
baths. As an example, Fig.~\ref{pronounced} shows the longitudinal
component of the central spin's oscillations for the bath with
small $\rho=5\times 10^{-5}$. To make the individual oscillations
noticeable in the figure, we also chose relatively small
$\omega=2$.

\begin{figure}
\vspace{-0.5cm}
\centerline{\includegraphics[width=95mm,angle=0,clip]{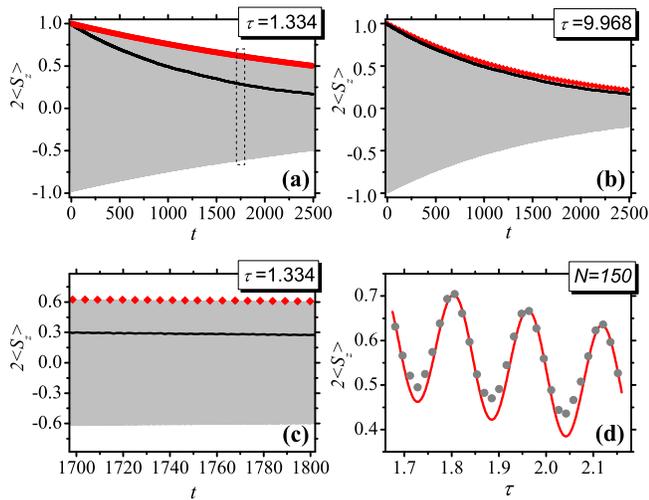}}
\vspace{-0.5cm} \caption{(Color online) Simulation results for the
fast bath with $\omega=20$ and $\rho=0.3$. Horizontal axes on
panels (a)--(c) represent the total time $\tilde t=4N\tau$.
Individual oscillations within the rotary echo signal (gray) have
very high frequency, and are not well resolved in the picture.
Analytical predictions for the rotary echo amplitudes (red
diamonds), as obtained from Eq.~\ref{moreacur}, match the
simulations very well. The envelope of the conventional Rabi
oscillations with the same parameters is shown with the black
solid line. (a) and (b): The short-$\tau$ and long-$\tau$ regimes,
respectively; the values of $\tau$ are chosen according to
Eq.~\ref{besttau}, the values of $\tau$ are shown in the graphs.
(c): zoom-in view of the marked area in (a), demonstrating good
agreement between analytics and numerics. (d): $\langle
S^z(N)\rangle$ as a function of $\tau$ for $N=150$. Gray dots are
the values obtained from simulations, and red line is obtained
analytically from Eq.~\ref{moreacur} (short-$\tau$ regime). }
\label{fastgraphs}
\end{figure}

Different regimes of the rotary echo decay for a slow bath are
demonstrated in Fig.~\ref{slowgraphs}. In agreement with our
analytical results, the rotary echo envelope survives much longer
than the ordinary Rabi oscillations, as long as the driving
reversal period is short, $\tau< |P|^{-1}$. As $\tau$ approaches
$|P|^{-1}$, the rotary echoes die off with virtually the same rate
as the Rabi oscillations; the numerically simulated decay is in
complete quantitative agreement with our analytical result
(\ref{resulttext}). Note that although $\rho=0.005$ may look like
a small number, but due to the long evolution time, the bath
dynamics is important, and the rotary echo decay differs from the
case of a purely static bath.

Figure \ref{fastgraphs} presents typical simulation results for a
fast bath. Again, one can see that the rotary echo protocol is
efficient in the short-$\tau$ regime. The numerical results show
that Eq.~(\ref{moreacur}) very accurately describes the echo
amplitudes. In particular, numerical simulations in
Fig.~\ref{fastgraphs}(d) demonstrate that performance of the
rotary echo protocol, in quantitative agreement with
Eq.~\ref{moreacur}, indeed depends on the value of $\tau$ in an
oscillatory fashion for fixed $N$, and the best protection against
decoherence is provided when $\tau$ is chosen according to
Eq.~\ref{besttau}. Also note that $\rho=0.3$, somewhat
non-intuitively, corresponds to a regime of the fast bath, in
spite of the fact that $R/b\sim 0.4<1$.

\section{Rotary echo protocol for realistic bath of N atoms in diamond}

\label{sec:realNV}

As mentioned above in Sec.~\ref{sec:description}, the  realistic
bath consisting of substitutional N atom (P1 centers) in diamond
has more complex structure than a single O-U field. Due to the
complex internal structure of the P1 centers, where an electron
spin $S=1/2$ is strongly ($\sim 100$~MHz) coupled to a nuclear
$^{14}$N spin $I=1$ via anisotropic hyperfine interaction, the
bath of P1 centers contains six different spectral groups
\cite{HansDobAwsch08,nvdd1,nvdd4,deLange}. Correspondingly, such a
bath should be described as a sum of six O-U noise fields
\cite{nvdd4} $B_k(t)$ ($k=1,\dots6)$, with the corresponding
parameters $b_k$ and $R_k$.

Besides, the NV electron spin interacts with the nuclear spin of
the NV's own $^{14}$N atom with the hyperfine constant
$A_0=-2\pi\times 2.16$ MHz, which has the same order of magnitude
as $b_k$, and has to be taken into account. The nuclear spin
state, and thus the hyperfine field created by it, is static on a
timescale of a single experimental run, but varies from one run to
another, and therefore can be described as a static field, which
randomly assumes the values $0, \pm A_0$ (corresponding to the
three states of the nuclear $^{14}$N spin $I=1$).

Below, we use the analytical results obtained above to investigate
the realistic bath of P1 centers, taking into account the multiple
noise fields and the on-site hyperfine coupling to the NV's own
$^{14}$N nuclear spin. We present analytical results, and compare
them with the direct numerical simulations, demonstrating the
quantitative agreement between analytics and numerics.

\subsection{Decoherence by multiple noise fields}

Because the equation of motion (\ref{stdiffeq}) is linear in $B$,
our theory can be generalized to include six noise fields.
Introducing independent oscillator modes for each of the noise
fields, $a_k$, and taking $b=\sqrt{\sum b^2_k}$ to define the
dimensionless quantities according to Eq.~\ref{dimless}, we find
that the evolution of the NV spin is governed by the dynamical
equation similar to Eq.~\ref{dyn}, but now with a modified
pseudo-Hamiltonian:
\begin{equation}
\label{dynsix} \hat{H}=\sum\limits_{k=1} ^6 \left[\rho_k\,
a^\dag_k a_k - \beta_k \frac{a_k+ a^\dag_k}
{\sqrt{2}}\hat{g}_z\right]- \omega\hat{g}_x,
\end{equation}
where $\rho_k= R_k/(\sqrt{2}b)$ and $\beta_k=b_k/b$.

It is seen that the presence of the external driving leads to
mixing between different oscillator modes, so that the six
oscillators are no longer independent. We can straightforwardly
extend the analysis described above (single noise with large
driving $\omega\gg1$) to the six-oscillator case, and readily
recover Eqs.~\ref{approxU}--\ref{hsmall}, with the simple
replacement of the combination, $(a+a^\dag)$ by $\sum \beta_k
(a_k+ a^\dag_k)$. Consequently, Eq.~\ref{SWres}, with
$\langle0_z|$ now denoting the state vector
$(0,0,1)\otimes\prod\langle0_k|$, yields correct result for the
rotary echo amplitude. However, the evaluation of $\langle
S^z(N)\rangle$ now becomes complicated, because now the reduced
pseudo-Hamiltonian
\begin{equation}
\label{hsmallmf} \hat{h}(\omega)= \sum_k \rho_k\, a^\dag_k a_k
-\omega\hat{s}_x - \frac{\left(\sum_k \beta_k (a_k+
a^\dag_k)\right)^2}{4\omega} \hat{s}_x,
\end{equation}
involves a mixture of different oscillator modes.

We can map the sum $\sum \beta_k (a_k+ a^\dag_k)$ onto a single
effective oscillator coordinate $c_1+c^\dag_1$, while keeping the
remaining five modes, $c_\mu$, $\mu=2,...,6$,  decoupled from the
spin. This is effected by a linear transformation of the original
oscillator operators via a real orthogonal $6\times6$ matrix
$\{\gamma_{kj}\}$, such that $\gamma_{k1}=\beta_k$. Introducing
$\bar{\rho}_k=\sum_i \rho_i\gamma_{ik}^2$ and $g_{jk}=\sum_i
\rho_i\gamma_{ij}\gamma_{ik}$, the transformed operator reads
\begin{equation}
\label{hsmallcc} \hat{h}(\omega)= \hat{h}_1(\omega)+
\hat{h}_5+\sum_{\mu=2}^6g_{1\mu}(c^\dag_1c_\mu+c^\dag_\mu c_1),
\end{equation}
where
\begin{eqnarray}
&&\hat{h}_1(\omega)= \bar{\rho}_1\, c^\dag_1
c_1- \frac{(c_1+c_1^\dag)^2}{4\omega} \hat{s}_x -\omega\hat{s}_x, \nonumber \\
&&\hat{h}_{5}=\sum_{\mu=2}^6\bar{\rho}_\mu c^\dag_\mu c_\mu
+\sum_{\mu>\nu=2}^6g_{\nu\mu}(c^\dag_\nu c_\mu +c^\dag_\mu
c_\nu).\nonumber
\end{eqnarray}
Here $\hat{h}_1$ replicates Eq.~\ref{hsmall}, with $c_1$
standing for the only oscillator mode coupled to the pseudo-spin,
and $\hat{h}_5$ is the internal pseudo-Hamiltonian of the
remaining five oscillator modes. The last term in Eq.
(\ref{hsmallcc}) is the interaction between $c_1$ and the
remaining five modes, characterized by the strength,
\begin{equation}
\label{lambda} g=\sqrt{\sum_{\mu=2}^6g_{1\mu}^2}=
\sqrt{ \frac12\sum_{l,k=1}^6( \rho_l-\rho_k)^2\beta_l^2\beta_k^2}.
\end{equation}

The central spin dynamics can be studied analytically in the
regime when this coupling is weak, i.e.\ when $g\ll |P_1|$ holds,
where $P_1=\sqrt{\bar{\rho}_1^2 -i\bar{\rho}_1/\omega}$ is the
typical level spacing of the Hamiltonian $\hat{h}_1$. Assuming
that there is no resonance between the lowest levels of
$\hat{h}_1$ and $\hat{h}_5$, the contribution of the interaction
Hamiltonian $h_5$ to the eigenvalues of $\hat{h}(\omega)$ is of
order of $\sim g^2/|P_1|$, and can be neglected.

Within this approximation, $\hat{h}_5$ decouples from $\hat{h}_1$,
and gives no contribution to the matrix element (\ref{SWres}). In
this case, the rotary echo decay is caused exclusively by the
$c_1$ mode, whose correlation decay rate is
$\bar{\rho}_1=\sum_k\rho_k \beta_k^2$. Hence the six O-U noise
fields are essentially combined into a single effective O-U field,
with the rms $b_e=\sqrt{\sum b^2_k}$, and the correlation decay
rate $R_e=\sum_kR_k b_k^2/b^2$. As a result, Eq.~\ref{resulttext}
and the subsequent results of Sec.~\ref{sec:decay} for $\langle
S^z(N)\rangle$ can be used, with replacement of $\rho$ by
$\sum_k\rho_k\beta_k^2$.

\begin{table}[b]
\caption{Characteristics of the six noise fields used in
numerical simulations, as measured in experiments \cite{deLange}. The
resulting effective
parameters are $R_e=42.33\,ms^{-1}$ and $b_e=3.59\,\mu s^{-1}$. }
\begin{tabular}{r |c |c| c| c| c| c}
\hline\hline
\# of field\,\, & 1 & 2 & 3 & 4 & 5 & 6 \\
\hline b\,($\mu s^{-1}$)\,\, &\,\, 0.83\,\, &\,\, 1.59\,\, &\,\,
1.63\,\, &\,\, 1.58\,\,
&\,\, 0.8 \,\,&\,\, 1.97\,\, \\
\hline
R\,($ms^{-1}$)\,\, & 39 & 42 & 139 & 7 & 4 & 6 \\
\hline
\end{tabular}
\label{table}
\end{table}

\begin{figure}
\vspace{-0.5cm}
\centerline{\includegraphics[width=95mm,angle=0,clip]{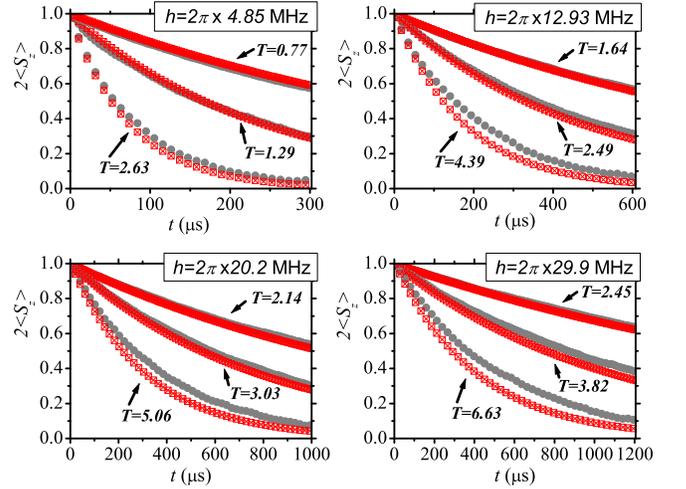}}
\vspace{-0.5cm} \caption{(Color online) Numerical simulations with
six noise fields. The rotary echo $\langle S^z(N)\rangle$ is
plotted as a function of total time $t=4NT$ for different values
of the quarter-periods $T$ (given in $\mu$s) of the driving
reversal. The four figures correspond to different driving
amplitudes, $h=2\pi\times 4.85$~MHz ($\omega=6$), $h= 2\pi\times
12.93$~MHz ($\omega=16$), $h= 2\pi\times 20.2$~MHz ($\omega=25$),
and $h= 2\pi\times 29.9$~MHz ($\omega=37$). The bath parameters
are given in Table \ref{table}. The gray dots denote the
numerically obtained $\langle S^z\rangle$, and the red symbols
correspond to the analytical results of Eq.~\ref{moreacur}, where
we assumed a single effective noise field with the parameters
$R_e$ and $b_e$, as explained in the main text.} \label{sixnoise}
\end{figure}

With the fixed values of $R_k$ and $B_k$, as measured in recent
experiments \cite{deLange}, the weak-coupling regime $g\ll |P_1|$
corresponds to a limited magnitude of $\omega$. At the same time,
in order to ensure good protection against decoherence, $\omega$
should not be small. The experimentally measured bath parameters
correspond to $\bar{\rho}_1\approx 0.00834$ and $g\approx 0.0143$,
which means that the weak-coupling regime holds for $\omega\ll
43$. Thus, we can expect that the analytical results above are
quantitatively accurate, say for $5\lesssim\omega\lesssim 20$, or,
in dimensional units, for an experimentally meaningful range of
$2\pi\times 4$~MHz~$\lesssim h\lesssim 2\pi\times 16$~MHz.

To gain quantitative insights, we performed numerical simulations,
starting from the rotating frame Hamiltonian,
$H=hS^x+\sum_kB_k(t)S^z$, with the experimentally measured
parameters of the six bath fields. In Fig.~\ref{sixnoise}, the
numerical results are compared with the analytical values of
$\langle S^z(N)\rangle$, obtained from Eq.~(\ref{moreacur}) with
$\rho=\bar{\rho}_1$. The values of $\tau$ are chosen according to
Eq.~(\ref{besttau}) to ensure the best decoupling from the bath.
Analytical and numerical results agree very well in the
experimentally relevant regime of short $\tau$.

For stronger driving, beyond the weak-coupling regime, six bath
fields start interacting with each other. In this regime, our
approximate analytical results predict faster rotary echo decay
than is actually seen from the direct simulations. Thus, the
analytics above can be considered as an estimate from below for
the real decay curve in this regime, and therefore remains useful
even for strong driving.

It is important to note that the effective values of $b_e$ and
$R_e$, which govern the rotary echo decay, are the same as those
which govern the free decay, spin echo dynamics, and the response
to dynamical decoupling of the NV center's spin \cite{nvdd1,
nvdd4, deLange}. Thus, the same picture of a single effective O-U
random field, with the same parameters, can describe a wide
variety of the dynamical regimes for the controlled spin of a NV
center.

\subsection{Influence of the hyperfine coupling}

In order to incorporate the on-site hyperfine coupling, the
pseudo-Hamiltonian (\ref{dyn}) should be modified in a
straightforward manner, acquiring the form
\begin{equation}
\label{dynHyp} \hat{H}=\rho a^\dag a - \omega\hat{s}_x - (a+
a^\dag+\lambda I^z_0) \frac{\hat{s}_z}{\sqrt{2}},
\end{equation}
where $\lambda=A_0/b$. Provided that $\lambda\ll\omega$, as it
happens in real experiments, further analysis can be performed
along the lines of Sec.~\ref{sec:decay} leading to the result akin
to Eq.~(\ref{SWres}). Namely, for $N/\omega^2\ll1$, the rotary
echo amplitude is well approximated by
\begin{equation}
\label{Sl} \langle
S^z_\lambda(N)\rangle\approx\frac16\sum_{I^z_0=0,\pm1}\langle
0_z|U_{\lambda I^z_0}^N|0_z\rangle,
\end{equation}
where $U_\lambda=e^{-\tau
\hat{h}_{\lambda +}}e^{-2\tau \hat{h}_{\lambda -}} e^{-\tau
\hat{h}_{\lambda +}}$ is determined by the reduced
pseudo-Hamiltonian operators,
\begin{equation}
\label{hsmHyp} \hat{h}_{\lambda \pm}= \rho\, a^\dag a
\mp\omega\hat{s}_x \mp\frac{(a+ a^\dag+\lambda)^2}{4\omega}
\hat{s}_x.
\end{equation}
Analytic form of $\langle 0_z|U_{\lambda}^N|0_z\rangle$ is found
in Appendix E. Because of its cumbersome form we do not present
the general analytical answer here. Instead we bring the
short-$\tau$ result,
\begin{equation}
\label{withHyp} 2\langle S^z_\lambda(N)\rangle\approx
\exp\!\left(-\frac{\rho\tau^3}{3\omega^2}N\right)\!\!\left[
\frac13 + \frac23
\exp\!\left(-\frac{\lambda^2\rho\tau^3}{3\omega^2}N\right)
\right]\!.
\end{equation}
This suggests that the hyperfine coupling leads to the suppression
of $2/3$ fraction of the total signal by
$\exp\bigl(-N\lambda^2\rho\tau^3/3\omega^2\bigr)$. With the
realistic experimental numbers we have $\lambda^2\approx 14.3$.
Therefore, the $2/3$ fraction of the signal decays much faster than
the remaining part.

Some comments on the approximation leading to the analytic form
for $\langle 0_z|U_\lambda^N|0_z\rangle$ and eventually to Eq.
(\ref{withHyp}) are appropriate. The approximation is base on the
assumption, $\lambda\ll\omega$. At the same time, its accuracy is
guaranteed only when $\psi>\lambda/\omega$, where $\psi$ has a
short-time asymptotics (\ref{psiasympsmall}) [cf. Eq.
(\ref{badconf}) in Appendix \ref{AppC}]. The latter condition can
be violated because of a small value of $\rho$ or ultra-short
$\tau$. By considering the static limit $\rho\to 0$ it can be
shown that the short-$\tau$ correction to $\langle
0_z|U_\lambda^N|0_z\rangle$ is negative and $\propto
(\lambda/\omega)^2 \sin^4\bigl(\frac12\tau
\sqrt{\omega^2+\lambda^2/2}\bigr)$. In contrast to the large-$N$
correction, Eq. (\ref{CorrText}), this correction does not
accumulate with $N$, and gets completely washed out for
$N\sim\omega$ and larger. Another difference is that while Eq.
(\ref{CorrText}) is due to the {\it dynamics} of the noise field,
this correction is because of {\it static} hyperfine and noise
fields. In order to get rid of this negative correction, one can
choose $\tau$ to satisfy $\sin\bigl(\frac12\tau
\sqrt{\omega^2+\lambda^2/2}\bigr)=0$, or $T\sqrt{h^2+A_0^2}=2\pi
k$ in dimensional units. \cite{AielHirCap12} Note that the obvious
conflict between this choice and the best decoupling condition
Eq.~(\ref{besttau}) is insignificant, and the latter condition can
be neglected. This is because the parameter domain where the
$\propto (\lambda/\omega)^2$ correction is tangible, the large-$N$
correction Eq. (\ref{CorrText}) is negligibly small.

\begin{figure}
\vspace{-0.5cm}
\centerline{\includegraphics[width=95mm,angle=0,clip]{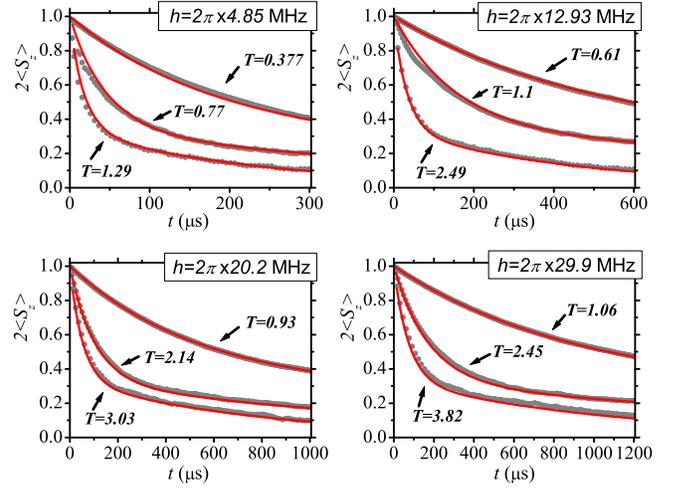}}
\vspace{-0.5cm} \caption{(Color online) Simulation results for the
rotary echo decay in the presence of six O-U noise fields and the
hyperfine interaction with $A_0=2\pi\times 2.16$ MHz. Parameters
of the noise fields are taken from Table \ref{table}. The four
panels correspond to four different driving amplitudes ($h/2\pi=
4.85$, $12.93$, $20.2$, and $29.9$~MHz, shown at the top of each
panel). Three curves in each panel represent the results obtained
for three different durations $T$ of the protocol quarter-periods,
the values of $T$ (in $\mu$s) are shown on the graphs. The
simulation results are denoted by gray dots, showing $\langle
S^z_\lambda(N)\rangle$ as a function of total time $t=4NT$. Red
symbols are the analytical results obtained from Eqs.~(\ref{Sl})
and (\ref{Ulans}).} \label{hypfine}
\end{figure}

In Fig.~\ref{hypfine} we compare our analytical predictions with
the results of numerical simulations. Simulations are based on the
rotating-frame Hamiltonian, $H=hS^x+
\left(\sum_kB_k(t)+A_0I^z_0\right)S^z$, where $I^z_0$ takes three
values, $\pm1$ and $0$, with equal probability. The analytical
results agree well with the numerical simulations, justifying the
approximations made in course of derivation. To highlight the
$\propto (\lambda/\omega)^2$ correction, we set
$T\sqrt{h^2+A_0^2}=4\pi$ and $16\pi$, for the plots with
$h=2\pi\times 4.85$~MHz, $T=0.377\,\mu$s and $h=2\pi\times
12.93$~MHz, $T=0.61\,\mu$s, respectively. The protocol
quarter-periods for the remaining plots are chosen to satisfy the
best decoupling condition Eq.~\ref{besttau}. Divergencies between
the simulated points and analytical curves for $h=2\pi\times
4.85$~MHz, $T=0.77\,\mu$s and $h=2\pi\times 12.93$~MHz,
$T=1.102\,\mu$s are noticeable for the short total times with
$N<\omega$. This demonstrates the onset of the $\propto
(\lambda/\omega)^2$ correction, in full agreement with the above
discussion.

\section{Conclusions}

\label{sec:discussion}

We considered the decay of rotary echoes for a central spin,
decohered by a surrounding environmental spins, focusing on the
symmetric rotary echo protocol which ensures good protection
against decoherence for strong driving. We approximated the impact
of the spin bath by a random time-varying magnetic field,
described as Ornstein-Uhlenbeck random process (which is
Markovian, Gaussian, and stationary). We obtained analytical
description for different dynamical regimes, and applied our
analytical results to investigating the decay of the rotary echoes
for a NV center in diamond decohered by a bath of substitutional N
atoms (P1 centers), which is the main decoherence source in type
Ib diamonds.

Our analytical treatment is based on the mapping of the problem on
the model of a spin $S=1$ coupled to a single bosonic mode, which
has been established in Sec.~\ref{sec:decay}. We note here an
interesting parallel with the Wiener-Hermite expansion, applied
earlier in the study of the Landau-Zener transition, when crossing
energy levels are subject to fluctuations due to a noisy
environment. \cite{Kayanuma} This expansion appears naturally in
our approach; in the present paper we did not pursue this
direction, but it might be an interesting subject for further
research.

Analytical form of the rotary echo amplitudes is obtained in the
limit of large driving. Comparing our analytical results with
direct numerical simulations, we found very good quantitative
agreement. Our analysis identified two main sources of the rotary
echo decay. One is the decay caused by the random field during the
periods of constant driving, and this channel is most effective at
not-too-long times ($N/\omega^2\ll 1$). The other is the decay
caused by a more delicate mechanism: during each reversal of the
driving, the effective quantization axis of the spin slightly
changes, but these changes accumulate with time, and may become
important when the total number of reversals is large
($N/\omega^2\sim 1$). We have also found that the latter channel
is least destructive when the reversal time $\tau$ is commensurate
with the driving period, see Eq.~\ref{besttau}.

For sufficiently frequent reversals (small $\tau$) the symmetric
rotary echo protocol ensures excellent protection of the spin. It
is interesting to compare it with the protection offered by the
pulse-based dynamical decoupling: for instance, for small $\tau$
and slow baths with $\rho\ll 1$, the echoes for the pulse-based
decoupling decay as $\exp[-\rho\tau^3N/3]$, while the rotary
echoes decay as $\exp[-\rho\tau^3 N/(3\omega^2)]$ (here we assumed
that the time $2\tau$ between two reversals is the same as the
inter-pulse delay $2\tau$ in the Carr-Purcell-Meiboom-Gill
protocol \cite{Slichter}). This comparison shows that the decay
rate is diminished by a large factor $\omega^2$, although achieved
due to constant application of strong driving to the central spin.
For realistic parameters\cite{nvdd4}, $b=3.6\,\mu s^{-1}$ and
$h=2\pi\times 8$~MHz, the rotary echo decay time survives longer
by a factor of about $\omega^2\sim 100$. In realistic experiments,
on such time scales the decay of the NV spins would be probably
dominated by other relaxation mechanisms.

It is also worth noticing that the symmetric protocol considered
here, with the cycle $(\tau|2\tau|\tau)$, is more efficient than
the asymmetric protocol $(2\tau|2\tau)$, where the driving is
switched after every time interval of $2\tau$. Most easily this
can be seen in the limit of static bath, $\rho\to 0$, where the
echo decay $\langle S^z(N)\rangle$ can be found exactly. For large
$\omega$ and $N/\omega^3\ll1$, the echo amplitude of the
asymmetric protocol is given by the integral
\begin{eqnarray}
\label{asymcycle} \langle S^z(N)\rangle= \frac12-\int
\frac{d\xi}{\sqrt{\pi}}
e^{-\xi^2}\sin^2\Bigl(\bigl[\omega+\xi^2/2\omega\bigr]\tau
\Bigr)&&\nonumber\\
\times\sin^2 \left(\frac{\sqrt{2}N}\omega
\sin\Bigl(\bigl[\omega+\xi^2/2\omega\bigr]\tau \Bigr)\right).&&
\end{eqnarray}
Obviously, the above integral is always positive and
$N$-dependent. This dependence is noticeable in experimentally
relevant regime, $N\sim\omega$, $\tau\lesssim\omega$, where the
integral takes values of order $1$, leading to the suppression of
signal. In contrast, corresponding integral for the symmetric
protocol is $\sim 1/\omega^2$, for arbitrary $N$ and $\tau$.

Representing the real bath of P1 centers as six independent O-U
random fields (corresponding to the six spectral lines of the P1
centers), we considered the decay of rotary echoes for a NV spin,
also taking into account the on-site hyperfine coupling. We found
that for the experimentally interesting region of large (but not
too large) drivings, the six bath fields can be replaced by a
single effective O-U random field. The parameters $b_e$ and $R_e$,
which describe this field and hence govern the decay of rotary
echoes, are the same as the parameters which govern the free
decay, spin echo dynamics, and the response to dynamical
decoupling of the NV center's spin. Thus, the same picture of a
single effective O-U random field, with the same parameters, can
describe a wide variety of the dynamical regimes for the
controlled spin of a NV center \cite{nvdd1,nvdd4,deLange}.

Combining all results together, we identified the regimes where
the symmetric rotary echo protocol most efficiently protects the
NV spin. Our results may provide useful guidelines for
understanding of the dynamics of a driven NV spin, and for
planning future experiments. It can also be applied to a variety
of other spins coupled to the dilute dipolar baths, such as donors
in silicon or magnetic ions in non-magnetic host crystals.
\cite{Pla, Morton, Morello, Bertaina}

\section{Acknowledgements}
We thank M. E. Raikh and L. Cywinski for interesting and useful
discussions. Work at the Ames Laboratory was supported by the U.S.
Department of Energy, Office of Science, Basic Energy Sciences,
Division of Materials Sciences and Engineering. The Ames Laboratory
is operated for the U.S. Department of Energy by Iowa State
University under contract No. DE-AC02-07CH11358.

\appendix

\section{}

In this Appendix we establish the relation (\ref{approxU}) for the
single-cycle evolution operator, valid for $\omega\gg1$. We begin
with transforming the pseudo-Hamiltonian (\ref{dyn}), to one that
commutes with the large driving, $\omega \hat{g}_x$, within a
certain accuracy. This accuracy is specified by the values of $N$
and $\tau$. We are going to discard the terms $\sim\tau/\omega^2$
and $N\tau/\omega^3$, presuming $\tau\ll \omega^2$, and
$N\tau\ll\omega^3$. Hence the accuracy should extend up to ${\cal
O}\bigl(1/\omega^3\bigr)$. At the first step, we apply the
transformation
\begin{equation}
\label{W0} W_0=\exp[\epsilon A], \quad A=(a+a^\dag)\hat{g}_y,
\end{equation}
where $\epsilon$ is assumed small. Neglecting the terms of
order $\epsilon^k$, $k\geq 4$, we write
\begin{eqnarray}
\label{0SWop} W_0\hat{H}W_0^{-1} \approx &&\hat{H}+
\epsilon[A,\hat{H}]+
\frac12\epsilon^2[A,[A,\hat{H}]]\nonumber\\
&&+\frac16\epsilon^3[A,[A,[A,\hat{H}]]].
\end{eqnarray}
To cancel the term containing $\hat{g}_z$ in $\hat{H}$, we choose
the parameter $\epsilon=1/\sqrt{2}\,\omega$. Then the last term in
Eq. (\ref{0SWop}) should be kept in order to achieve the necessary
accuracy, since $\hat{H}$ contains a term of order of $\omega$. We
find:
\begin{widetext}
\begin{equation}
\label{0SW} W_0\hat{H}W_0^{-1}=\rho\, a^\dag a - \omega\hat{g}_x -
\frac1{2\omega} \hat{g}_x\left(\!\frac{a+
a^\dag}{\sqrt{2}}\right)^2 +\frac\rho\omega \hat{g}_y\frac{a-
a^\dag}{\sqrt{2}} -\frac{\rho}{2\omega^2} \hat{g}_y^2
+\frac1{3\,\omega^2} \hat{g}_z \left(\!\frac{a+
a^\dag}{\sqrt{2}}\!\right)^3\! + {\cal O}\bigl(1/\omega^3\bigr).
\end{equation}
To get rid of the leading inconvenient term, which does not
commute with $\hat{g}_x$, we next apply a transformation with the
operator,
\begin{equation}
\label{W1} W_1=\exp\left[\frac\rho {\sqrt{2}\,\omega^2}
(a-a^\dag)\hat{g}_z \right].
\end{equation}
This leads to the relation,
\begin{equation}
\label{1SW} W_1W_0\hat{H}W_0^{-1}W_1^{-1}=\rho\, a^\dag a -
\omega\hat{g}_x - \frac1{2\omega} \hat{g}_x\left(\!\frac{a+
a^\dag}{\sqrt{2}}\!\right)^2  -\frac{\rho}{2\omega^2} \hat{g}_y^2
+\frac1{3\,\omega^2} \hat{g}_z \left(\!\frac{a+
a^\dag}{\sqrt{2}}\!\right)^3\! +\frac{\rho^2} {\omega^2}
\hat{g}_z\frac{a+ a^\dag}{\sqrt{2}} + {\cal
O}\bigl(1/\omega^3\bigr).
\end{equation}
Subsequent transformations with
\begin{equation}
\label{W23} W_2=\exp\left[-\frac\rho {8\omega^3}
(\hat{g}_y\hat{g}_z +\hat{g}_z\hat{g}_y)\right],\qquad
W_3=\exp\left[-\frac{\hat{g}_y}{\sqrt{2}\,\omega^3}\left(\rho^2(a+a^\dag)
+\frac16 (a+a^\dag)^3\right) \right],
\end{equation}
discard further inconvenient terms, up to the third
order in $1/\omega$. As a result, we get
\begin{equation}
\label{smSW}\hat{H}=W^{-1} \hat{h}\, W + {\cal
O}(1/\omega^3),\qquad \hat{h}(\omega)=\rho\, a^\dag a -
\omega\hat{g}_x - \frac{(a+ a^\dag)^2}{4\omega} \hat{g}_x
+\frac\rho{2\omega^2}\left[{\bf 1}+ \frac12\hat{g}_x^2 \right],
\end{equation}
where $W=W_3W_2W_1W_0$ includes the above four subsequent
transformations in the mixed pseudospin-oscillator space. Some
comments on these transformations are in order. The first of them,
$W_0$, is unitary. $W_1$ is smaller than $W_0$ and invariant with
respect to the sign reversal of $\omega$. $W_2$ and $W_3$ contain
even smaller exponents, $\sim 1/\omega^3$, thus seem to be beyond
our accuracy goal. At this point one should be more careful, as
$\hat{h}$ still contains the large term $\omega\hat{g}_x$, which may
amplify the effect of $W_2$ and $W_3$. However, we will see that
this is not the case and the latter operators can be safely ignored.
From Eq. (\ref{SviaME}) we have
\begin{equation}
\label{afterSW}2\langle S^z(N)\rangle=\langle0_z| W^{-1}\left[
e^{-\tau \hat{h}_{+}}Ye^{-2\tau \hat{h}_{-}}Y^{-1}e^{-\tau
\hat{h}_{+}}\right]^N W|0_z\rangle +{\cal O}(N/\omega^3),
\end{equation}
with $\hat{h}_{\pm}=\hat{h}(\pm\omega)$ and
$Y=W(\omega)W^{-1}(-\omega)$. Here we have taken into account that
the $\tau$-dependence of the remainder of Eq. (\ref{afterSW}) is
inessential.
The large terms $\mp\omega\hat{g}_x$ commute with the
remaining parts of $\hat{h}_{\pm}$, so that we can separate
the corresponding exponents:
\begin{equation}
\label{sepoff} e^{-\tau \hat{h}_{\pm}}=e^{-\tau \hat{h}'_{\pm}}
e^{\pm\tau \omega\hat{g}_x}= e^{\pm\tau \omega\hat{g}_x}e^{-\tau
\hat{h}'_{\pm}}, \qquad \hat{h}'_{\pm}=\rho\, a^\dag a \mp
\frac{(a+ a^\dag)^2}{4\omega} \hat{g}_x
+\frac\rho{2\omega^2}\left[{\bf 1}+ \frac12\hat{g}_x^2 \right].
\end{equation}
Therefore the matrix element (\ref{afterSW}) is equal to
\begin{equation}
\label{smallh}\langle0_z| W^{-1}\left[ e^{-\tau
\hat{h}'_{+}}\bigl(e^{\tau \omega\hat{g}_x}Y e^{-\tau
\omega\hat{g}_x}\bigr) e^{-2\tau \hat{h}'_{-}}\bigl(e^{-\tau
\omega\hat{g}_x}Y^{-1}e^{\tau \omega\hat{g}_x}\bigr)e^{-\tau
\hat{h}'_{+}}\right]^N W|0_z\rangle.
\end{equation}
\end{widetext}
Now consider the product, $e^{\tau\omega\hat{g}_x}
Ye^{-\tau\omega\hat{g}_x}$. Using the fact that $W_1$ is even with
respect to the sign of $\omega$ while the rest of $W_i$ are odd,
we write
\begin{equation}
\label{Vprod} e^{\tau\omega\hat{g}_x}
Ye^{-\tau\omega\hat{g}_x}=\tilde{W}_3 \tilde{W}_2 \tilde{W}_1
\tilde{W}_0^2\tilde{W}_1^{-1}\tilde{W}_2 \tilde{W}_3,
\end{equation}
where $\tilde{W}_i=e^{\tau\omega\hat{g}_x}
W_ie^{-\tau\omega\hat{g}_x}$ are given by Eqs. (\ref{W0}),
(\ref{W1}), and (\ref{W23}), upon the substitution,
$\hat{g}_{y,z}\to \hat{\tilde{g}}_{y,z}= e^{\tau\omega\hat{g}_x}
\hat{g}_{y,z}e^{-\tau\omega\hat{g}_x}$. Explicitly,
$\hat{\tilde{g}}_y=\cos(\tau\omega)\hat{g}_y+
\sin(\tau\omega)\hat{g}_z$, while
$\hat{\tilde{g}}_z=\cos(\tau\omega)\hat{g}_z-
\sin(\tau\omega)\hat{g}_y$. Thus we see that the large parameter
$\omega$ ends up in the arguments of sines and cosines, leaving
the transformations $\tilde{W}_2$ and $\tilde{W}_3$ with exponents
still suppressed as $1/\omega^3$. Exactly the same applies to the
other product, $e^{-\tau\omega\hat{g}_x}
Y^{-1}e^{\tau\omega\hat{g}_x}$, and the operators,
$e^{-\tau\omega\hat{g}_x} W_{2,\,3}e^{\tau\omega\hat{g}_x}$,
entering this product. Because there is no longer large parameter
remaining in $\hat{h}'_{\pm}$, these transformations are
negligible in Eq. (\ref{smallh}). Therefore, all $W_2$ and $W_3$
operators inside $W$ and $Y$ in Eq. (\ref{afterSW}) are negligible,
too. In turn, the two properties of $W_1$, namely that
$W_1W_0^2W_1^{-1}=W_0^2+{\cal O}(1/\omega^3)$, and $W_1
|0_z\rangle= |0_z\rangle$, allow us completely discard $W_1$ from
equation (\ref{afterSW}) as well. As a result, we arrive at the
relation,
\begin{equation}
\label{afterSWsimp} 2\langle S^z(N)\rangle=\langle0_z| W^{\dag
}_0\left(\bar{U}^N \right)W_0|0_z\rangle+ {\cal O}(N/\omega^3),
\end{equation}
with
\begin{equation}
\label{UbarOnly} \bar{U}= e^{-\tau \hat{h}_{+}}W^2_0e^{-2\tau
\hat{h}_{-}}W^{\dag\, 2}_0e^{-\tau \hat{h}_{+}},
\end{equation}
which is a more disclosed form of Eq. (\ref{approxU}). Obviously,
the difference between $W_0|0_z\rangle$ and $|0_z\rangle$ is
$\sim1/\omega$. In addition, the norm of $\bar{U}^N$ is less than
$1$. Hence the contribution from the $W_0$-operators outside the
parentheses in the matrix element Eq. (\ref{afterSWsimp})
is at most \cite{footnote} $\sim 1/\omega$ and does not accumulate
with $N$. We neglect this contribution and write:
\begin{equation}
\label{viaUbar} \langle S^z(N)\rangle= \frac12\langle0_z|\bar{U}^N
|0_z\rangle.
\end{equation}
Using the smallness of the exponent of $W_0$, we further separate the
principal part of $\bar{U}$ and consider $W_0$ in Eq.
(\ref{UbarOnly}) as a correction. This is done by introducing the
operator
\begin{equation}
\label{Kpm} \Xi(\tau)=\sqrt{2}\,e^{\tau \hat{h}_{+}}\left[(a+
a^\dag) \hat{g}_y\right]e^{-\tau \hat{h}_{+}}.
\end{equation}
To the first order in $1/\omega$, Eq. (\ref{UbarOnly}) reads
\begin{equation}
\label{corr2}\bar{U}= U_0+ \frac{1}\omega\, U_1,
\end{equation}
where the principal part of the single cycle time evolution
operator is given by
\begin{equation}
\label{princrpal}U_0=\exp[-\tau \hat{h}_{+}] \exp[-2\tau
\hat{h}_{-}] \exp[-\tau \hat{h}_{+}],
\end{equation}
whereas
\begin{equation}
\label{perturb} U_1= U_0 \Xi(\tau) -\Xi(-\tau)U_0
\end{equation}
is a perturbation. In Appendix C we demonstrate that the last term
in Eq. (\ref{corr2}) yields in fact a correction $\sim 1/\omega^2$
to the eigenvalue of $\bar{U}$. Therefore the rotary echo amplitudes are
well described by Eq. (\ref{SWres}), provided
that $N/\omega^2\ll1$. For larger $N$, however, the contribution of $U_1$
is considerable. In Appendix D we calculate this contribution
for larger numbers, $N/\omega^2\lesssim1$.

\section{}

\label{AppB}

In this Appendix we evaluate $\langle S^z(N)\rangle$ from Eq.
(\ref{SWres}), by solving the eigenvalue problem for $U_0$.

The operators $\hat{h}_{\pm}$ contain $\hat{g}_x$, so that we
introduce the normalized eigenvectors, $\hat{g}_x|X_0\rangle=0$,
$\hat{g}_x|X_{\pm}\rangle=\pm i|X_{\pm}\rangle$. In the basis of
Eq. (\ref{O3}) their explicit forms are:
\begin{equation}
\label{eigenvecsx}|X_0\rangle= \left(\!\begin{array}{c}
1\\
0\\
0
\end{array}\!\right), \quad |X_{\pm}\rangle=
\frac1{\sqrt{2}} \left(\!\begin{array}{c}
0\\
\pm i\\
1
\end{array}\!\right).
\end{equation}
Hence the state $|0_z\rangle$ is
\begin{equation}
\label{03viaXpm} |0_z\rangle=
\frac1{\sqrt{2}}\Bigl(|X_+\rangle+|X_-\rangle\Bigr)\otimes
|0\rangle,
\end{equation}
where $|0\rangle$ is the ground state of the oscillator mode.

The last term of $\hat{h}$, Eq. (\ref{smSW}), is a scalar in the
subspace $|X_{\pm}\rangle$, decoupled from the oscillator mode.
This term leads to longitudinal relaxation along the $x$ axis.
Although it is the easiest to handle, throughout this Appendix we
ignore it, as its contribution for $N/\omega^2\ll1$ is negligible.
Then for $U_0$ we write
\begin{equation}
\label{U0viaV} U_0=V_0 |X_0\rangle\langle X_0|+
V_{+}|X_+\rangle\langle X_+|+ V_-|X_-\rangle\langle X_-|,
\end{equation}
where $V_0=\exp(-4\tau\rho\, a^\dag a)$, and
$V_{\pm}=V(\pm\omega)$ with
\begin{widetext}
\begin{equation}
\label{Vpm}
V(\omega)=V_{\frac12}(-\omega)V_{\frac12}(\omega),\quad
V_{\frac12}(\omega)=\exp\left(-\tau\rho a^\dag a
-i\frac{\tau}{4\omega}(a+a^\dag)^2\right) \exp\left(-\tau\rho
a^\dag a +i\frac{\tau}{4\omega}(a+a^\dag)^2\right).
\end{equation}
\end{widetext}
From Eqs. (\ref{eigenvecsx})- (\ref{U0viaV}) we express the matrix
element as
\begin{equation}
\label{me1}\langle0_z| U_0^N |0_z\rangle= \frac12\langle0| V_{+}^N
+ V_{-}^N |0\rangle=\text{Re}\langle0| V^N(\omega) |0\rangle.
\end{equation}
To calculate the latter, we bring $V(\omega)$ into the form of a
single exponent. This can be done by noticing that the
combinations,
\begin{equation}
\label{su2} K_x=i\frac{a^{\dag\,2}+a^2} 2, \quad K_y=
\frac{a^{\dag\,2}-a^2}2, \quad K_z=a^\dag a+ \frac 12,
\end{equation}
satisfy the commutation relations of the $su(2)$ Lie algebra,
$[K_\alpha,K_\beta]=2i\,\epsilon_{\alpha\beta\gamma}\,K_\gamma$.
This algebraic property allows to write
$V_{\frac12}$ in terms of the single exponent,
\begin{equation}
\label{halfcycle}
V_{\frac12}(\omega)=\exp[-\phi(\vec{n}\vec{K})+\tau\rho],
\end{equation}
where the real positive $\phi$ and the complex unit vector
$\vec{n}$ are given in terms of the combination, $P=\sqrt{\rho^2
-i\rho/\omega}$, and its complex conjugate, as follows:
\begin{eqnarray}
\label{hcn}&&\cosh\phi=\cosh \tau P\cosh\tau
P^*+\frac{\rho^2}{|P|^2}\sinh \tau P\sinh\tau
P^*,\nonumber \\
&&\,\nonumber\\
&&n_x= \frac{\sinh\tau P\cosh\tau P^*}{2\omega P\sinh\phi}
-\frac{\sinh\tau P^*\cosh\tau P}{2\omega P^*\sinh\phi},\nonumber \\
&&n_y=i\rho\,\frac{\sinh\tau P\sinh\tau
P^*}{\omega|P|^2\sinh\phi}, \\
&&n_z=in_x+\rho\frac{\sinh\tau P\cosh\tau P^*}{ P\sinh\phi}+
\rho\frac{\sinh\tau P^*\cosh\tau P}{ P^*\sinh\phi}. \nonumber
\end{eqnarray}
In the same way we represent $V(\omega)$ as
\begin{equation}
\label{fullcycle}
V(\omega)=\exp[-\psi(\vec{q}\,\vec{K})+2\tau\rho],
\end{equation}
with
\begin{eqnarray}
\label{fcm1}&&\cosh\psi=\cosh^2\phi
+(1-2n_y^2)\sinh^2\phi,\\
&&q_x=\frac{n_x\cosh\phi+in_yn_z\sinh\phi}{\sqrt{(1-n_y^2)
(\cosh^2\phi-n_y^2\sinh^2\phi)}},\\
\label{fcm2}
&&q_y=0, \nonumber \\
&&q_z=\frac{n_z\cosh\phi-in_xn_y\sinh\phi}{\sqrt{(1-n_y^2)
(\cosh^2\phi-n_y^2\sinh^2\phi)}}.
\label{fcm}
\end{eqnarray}
Note that while the angle $\psi$ is a real number not sensitive
to the sign of $\omega$, the complex unit vector $\vec{q}$ depends
on that sign, reflecting the difference between $V_+=V(\omega)$
and $V_-=V^*(\omega)$.

The eigenvalue problem of $(\vec{q}\,\vec{K})$ with a general (complex)
$\vec{q}$ is solved by employing the identity,
\begin{equation}
\label{Blt} (\vec{q}\,\vec{K})=e^{-i\alpha
K_y}e^{-i\beta(K_z-iK_x)} K_z e^{i\beta(K_z-iK_x)} e^{i\alpha
K_y},
\end{equation}
where $e^{-2i\alpha}=q_z-iq_x$ and $2\beta=\frac{q_y}{q_z-iq_x}$.
Hence the eigenfunctions of $(\vec{q}\,\vec{K})$ are
$|\chi_n\rangle = e^{-i\alpha K_y}e^{-i\beta(K_z-iK_x)}
|n\rangle$, where $|n\rangle$ are the oscillator eigenstates
(eigenfunctions of $K_z$), and the corresponding eigenvalues are
equal to $(n+1/2)$. In our case we have $\beta=0$. At the same
time, up to an inessential factor, acting on a state function with
the squeezing operator, $e^{-i\alpha K_y}$, results in the
multiplication of its argument by $e^{i\alpha}$. We thus get
$(\vec{q}\,\vec{K})\chi_n(\xi)= (n+1/2) \chi_n(\xi)$, with
$n=0,1,2,...$, and the normalized eigenfunctions,
\begin{equation}
\label{eigfunc} \chi_n(\xi)=\left(\frac\nu\pi\right)^{\frac14}
\frac1{\sqrt{2^nn!}}e^{-\nu\xi^2/2}H_n(\sqrt{\nu}\xi),
\end{equation}
where $\nu=q_z+iq_x$ and $H_n$ are the Hermite polynomials.
Explicitly, we have solved the eigenvalue problem,
\begin{equation}
\label{eigvalprob} V(\omega)|\chi_n\rangle=
\exp\!\bigl[-\psi(n+1/2)+2\tau\rho\bigr]\! |\chi_n\rangle.
\end{equation}
On the ground of this solution we find the matrix element Eq.
(\ref{me1}) as follows. The normalized wavefunction,
$|0\rangle=\pi^{-\frac14} \exp(-\xi^2/2)$, is expanded in terms of
the eigenstates of $V(\omega)$ by $|0\rangle=\sum_{n\geq0}
C_n|\chi_n\rangle$, with the coefficients,
\begin{equation}
\label{cn}C_n=\frac{\nu^{\frac14}}{\sqrt{\pi2^nn!}}\int d\xi
e^{-\frac{\xi^2}2-\nu\frac{\xi^2}2}H_n(\sqrt{\nu}\xi).
\end{equation}
After the integration, survive only coefficients with even $n=2k$,
\begin{equation}
\label{cnexp}C_{2k}=\frac1{\nu^{\frac14}k!}\sqrt{\frac{(2k)!}{2^{2k-1}}}
\frac{(1-q_z+iq_x)^k}{(1+q_z-iq_x)^{k+\frac12}}.
\end{equation}
From this expansion and Eq. (\ref{eigvalprob}) we get:
\begin{eqnarray}
\label{me2} \langle0| V(\omega)^N
|0\rangle=\!\!\!&&e^{2N\rho\tau}\,\sum\limits_{k\geq0} C_{2k}^2
e^{-N\psi(2k+1/2)}\nonumber\\
=\!\!\!&&\frac{e^{2N\rho\tau}}{\sqrt{\cosh N\psi+ q_z\sinh
N\psi}},\qquad
\end{eqnarray}
where the last equality is due to the relation,
$\sum_{k\geq0}\bigl[(2k)!/2^{2k}(k!)^2\bigr]x^{2k}
=(1-x^2)^{-1/2}$.
It is straightforward to check that Eq.~(\ref{Qbig}) is simply a
different form of the relation Eq.~(\ref{fcm}) for $q_z$. Finally,
the result Eq.~(\ref{resulttext}) follows from  Eq. (\ref{me1}),
by taking real part of Eq. (\ref{me2}).

\section{}

\label{AppC}

In this Appendix we analyze the perturbation $U_1$, Eq.~(\ref{corr2}),
and show that its contribution to the eigenvalues of
$\bar{U}$ is basically $\sim1/\omega^2$.

From Eq. (\ref{Kpm}) it is straightforward to check that $\Xi
(\tau)$ does not have diagonal matrix elements in the eigenvector
basis of $U_0$. Therefore the perturbation $U_1$ does not have
such diagonal matrix elements either. In particular, as follows
from Eq. (\ref{perturb}), the only non-zero matrix elements of
$U_1$ are the ones between the states
$|\chi_n\rangle_{\pm}=X_{\pm}\otimes|\chi_n^{\pm}\rangle$ and
$|m\rangle_0=X_0\otimes |m\rangle$, provided that $m=n+1+2j$,

$j=0, \pm1, \pm2, ...$. This means that to the first order in
$1/\omega$ there is no corrections to eigenvalues of $\bar{U}$ due
to $U_1$, and the spectrum of $\bar{U}$ is affected by $U_1$ only
to the order of $1/\omega^2$.

This proves our statement for general values of parameters.
However, some complications can arise when the eigenstates
$|\chi_n\rangle_{\pm}$ and $|m\rangle_0$ are degenerate. In the
remainder of this Appendix we demonstrate that the degeneracy does
not lead to any significant effect, as the resulting correction
from $U_1$ becomes comparable with the principal part from $U_0$
only when both these quantities are strongly suppressed, i.e.,
when the central spin is completely decohered.

The above mentioned degeneracy means that $V_{\pm}$ and $V_0$ have
equal eigenvalues. From the spectra of these operators found in
Appendix B one can see that this is the case when
\begin{widetext}
\begin{equation}
\label{degencon}-\psi(n+1/2)+2\tau\rho= -4\tau\rho m, \qquad
m=n+1+2j,\quad j=0,1,2,...,
\end{equation}
where $j$ is non-negative because $\psi>4\tau\rho$. Once
this happens, $U_1$ lifts the degeneracy, contributing $\sim 1/\omega$
in the eigenvalue of $\bar{U}$, and mixing the states
$|\chi_n\rangle_{\pm}$ with $|m\rangle_0$. More specifically, if
$_0\langle m|U_1|\chi_n\rangle_+=U_{1;m,n}$ and
$_+\langle\chi_n|U_1|m\rangle_0\equiv U_{1;n,m}$ denote matrix
elements between the degenerate states, the correct eigenvectors of
$\bar{U}$ are
\begin{equation}
\label{correigvec} \sqrt\frac{U_{1;n,m}}{U_{1;n,m}+U_{1;m,n}}
\,|\chi_n\rangle_+ \pm \sqrt\frac{U_{1;m,n}}{U_{1;n,m}+U_{1;m,n}}
\,|m\rangle_0,
\end{equation}
instead of $|\chi_n\rangle_+$ and $|m\rangle_0$, with the
corresponding eigenvalues,
\begin{equation}
\label{correigval}e^{-(n+1/2)\psi+ 2\tau\rho} \pm\frac1\omega
\sqrt{U_{1;n,m}U_{1;m,n}}= e^{-4\tau\rho\, m} \pm\frac1\omega
\sqrt{U_{1;n,m}U_{1;m,n}}.
\end{equation}
From more detailed analysis of Eq. (\ref{perturb}) we find that
the conjugate matrix elements of $U_1$ are the same,
\begin{equation}
\label{matel}_0\langle m|U_1|\chi_n\rangle_{\pm}=
_{\pm}\langle\chi_n|U_1|m\rangle_0\equiv U_{mn},
\end{equation}
leading to the corrected eigenvectors $(|\chi_n\rangle_+ \pm
|m\rangle_0)/\sqrt{2}$ and eigenvalues $e^{-(n+1/2)\psi+
2\tau\rho} \pm U_{mn}/\omega$.

Quantitatively, we can take the degeneracy in the $n$th channel
into account by modifying the term, $2k=n$, in the sum Eq.
(\ref{me2}). As $|m\rangle_0$ is orthogonal to $|0_z\rangle$, the
degenerate channel will contribute by the amount,
\begin{equation}
\label{degchan0}\frac{C_n^2}2\left[\left(e^{-(n+\frac12)\psi+
2\tau\rho}+ \frac{U_{mn}}\omega \right)^N +
\left(e^{-(n+\frac12)\psi- 2\tau\rho}- \frac{U_{mn}}\omega
\right)^N\right].
\end{equation}
Certainly, the only channels of interest are ones with the lowest
numbers, $n$ and $m$, or otherwise the corresponding eigenstates
do not contribute appreciably. We thus consider only
$e^{-(n+\frac12)\psi- 2\tau\rho}\sim 1$. From the definition it
also follows that $|U_{mn}|<1$, so that we have
\begin{equation}
\label{atthedeg}|U_{mn}|e^{(n+\frac12)\psi-
2\tau\rho}=|U_{mn}|e^{4m\tau\rho}\sim 1.
\end{equation}
Then the contribution Eq. (\ref{degchan0})
can be cast to the form
\begin{equation}
\label{degchan}
C_n^2e^{-N(n+\frac12)\psi+ 2N\tau\rho}\cosh\!\left[\frac
N\omega U_{mn}e^{(n+\frac12)\psi- 2\tau\rho}\right],
\end{equation}
so that the leading correction to Eq.
(\ref{resulttext}) is given by
\begin{equation}
\label{leadcorr}\sum\limits_{\text{degenerate}\atop {n, m}
}C_n^2e^{-4Nm\tau\rho}\left(\cosh\!\left[\frac N\omega
U_{mn}e^{4m\tau\rho}\right]-1\right).
\end{equation}
For small $N\ll\omega$ this correction is $\sim (N/\omega\bigl)^2$
and thus irrelevant. For larger $N$, on the other hand, the
exponent $e^{-4Nm\tau\rho}=e^{-N(n+\frac12)\psi+ 2N\tau\rho}$
kicks in, and the correction is suppressed exponentially, unless
$4m\tau\rho<1/\omega$ and $\psi<1/\omega$. Hence we conclude that
the decay of $\langle S^z(N)\rangle$ may appreciably deviate from
Eq. (\ref{resulttext}) only at small $\psi\ll1$.

Besides, on general grounds one should expect that the
perturbative arguments fail when the level spacing of the
non-perturbed operator $U_0$ is comparable with eigenvalues of the
perturbation. This can happen when
\begin{equation}
\label{badconf}\psi\lesssim1/\omega.
\end{equation}
We will now discuss the domain $\psi\ll1$ and show that the
correction is of order $\psi/\omega$. To treat the slow and fast
baths in equal mode, we explicitly describe the boundary of the
domain as
\begin{equation}
\label{smallpsi} \tau\rho\ll1,\quad \rho\tau^2/\omega\ll1.
\end{equation}
While for fast bath the first condition implies that the second
one fulfills, for slow bath the domain boundary is defined by the
second inequality, entailing the first one. To the leading order
over the small parameters $\eta=\tau\rho$, $\epsilon=\rho
\tau^2/\omega$, we arrive at the decomposition
\begin{eqnarray}
\label{decomp1} \exp(-\tau\hat{h})\!\!\!&&= \exp\!\left(\tau\omega
\hat{g}_x+\!\left[\frac\tau{\omega}
\frac{\hat{g}_x}2+\eta\left(\frac\tau\omega\right)^2
\frac{\hat{g}_x^2}6\right]\!\frac{(a+ a^\dag)^2}2\right)
\exp\!\left(\left[-\eta +\frac23\eta\epsilon\hat{g}_x \right]\!
a^\dag a\!\right) \exp\!\left(
\epsilon\frac{\hat{g}_x}2\frac{a^{\dag\,2} -a^2}2\right)\\
&&= \exp\!\left(\!-\epsilon\frac{\hat{g}_x}2\frac{a^{\dag\,2}
-a^2}2\right)\exp\!\left(\left[-\eta +\frac23\eta\epsilon\hat{g}_x
\right]\! a^\dag a\!\right)\exp\!\left(\tau\omega \hat{g}_x+
\!\left[\frac\tau{\omega}
\frac{\hat{g}_x}2+\eta\left(\frac\tau\omega\right)^2
\frac{\hat{g}_x^2}6\right]\!\frac{(a+ a^\dag)^2}2\right)\!.\quad
\label{decomp2}
\end{eqnarray}
Here and in subsequent calculations it is important to keep
operator content of the exponents, while we expand their scalar
coefficients. For this reason we still keep the terms
$\propto\epsilon\eta$. For later convenience we introduce the
parameter $\lambda=\rho\tau^3/4\omega^2$, though it is expressed
via $\eta$ and $\epsilon$. This choice descends from the fact that
while for fast bath $\lambda\ll1$, for slow bath it can be large
despite the restrictions Eq. (\ref{smallpsi}). Utilizing the
decomposition Eq. (\ref{decomp1}) with $\omega\to-\omega$ for
$\hat{h}_{-}$ and Eq. (\ref{decomp2}) for $\hat{h}_+$, we write
the half-cycle evolution operator $\hat{V}_{\frac12}(\omega)=
e^{-\tau \hat{h}_{+}}W^2_0e^{-\tau \hat{h}_{-}}$ in the form
\begin{equation}
\label{hatV} \hat{V}_{\frac12}(\omega)= \exp(-\epsilon
Q_1\hat{g}_x) \exp\!\left(\left[-\eta
+\frac23\eta\epsilon\hat{g}_x \right]\! a^\dag a\!\right)
\hat{A}(\omega) \exp\!\left(\left[-\eta
-\frac23\eta\epsilon\hat{g}_x \right]\! a^\dag
a\!\right)\exp(-\epsilon Q_1\hat{g}_x),
\end{equation}
where we introduced $Q_0=(a+ a^\dag)^2/3$, $Q_1=(a^{\dag\,2}
-a^2)/4$, and
\begin{equation}
\label{Adef} \hat{A}(\omega)=\exp\left(\!\hat{g}_x
\left[\!\omega+\frac{3Q_0}{4\omega} \!\right]\!\tau\right)
\exp(\lambda Q_0\hat{g}_x^2)
\exp\!\!\left(\frac{\sqrt{2}}\omega(a+
a^\dag)\hat{g}_y\!\right)\exp(\lambda Q_0\hat{g}_x^2)
\exp\left(\!-\hat{g}_x \left[\!\omega+\frac{3Q_0}{4\omega}
\!\right]\!\tau\right).
\end{equation}
First we analyze the domain of large $\lambda\gtrsim1$. We express
$\hat{A}(\omega)$ in terms of a single exponent. The three
exponents at the middle of product Eq. (\ref{Adef}) commute in the
oscillator sector, so that we have a combination,
$\exp(\alpha\hat{g}_x^2) \exp(\beta\hat{g}_y)\exp(\alpha
\hat{g}_x^2)$ with commuting $\alpha$ and $\beta$. It is easy to
see that $e_2=(0,1,0)$ is the eigenvector of this combination,
with the "eigenvalue" $\exp(-2\alpha)$. In the complementary
subspace spanned over $e_1=(1,0,0)$ and $e_3=(0,0,1)$, on the
other hand, the action of the pseudospin operators are expressed
via Pauli matrices $\sigma_i$ as $\hat{g}_x^2=(\sigma_z-1)/2$,
$\hat{g}_y=i\sigma_y$. Exploiting the algebra of Pauli matrices we
get:
\begin{equation}
\label{Awithneighb} \hat{A}(\omega)=\exp\left(\!\hat{g}_x
\left[\!\omega+\frac{3Q_0}{4\omega}
\!\right]\!\tau\right)\exp\left(2\lambda
Q_0\hat{g}_x^2+\frac{\sqrt{2}}\omega \frac{\lambda Q_0(a+a^\dag)}
{\sinh\lambda Q_0}\hat{g}_y \right)\exp\left(\!-\hat{g}_x
\left[\!\omega+\frac{3Q_0}{4\omega} \!\right]\!\tau\right).
\end{equation}
The same as above applies to this product; the three exponents
commute in the oscillator sector, so that after rotating around
$\hat{g}_x$ we get a single exponent,
\begin{equation}
\label{Asingexp}\hat{A}(\omega)= \exp\!\!\left(\!2\lambda
Q_0\hat{g}_x^2+ \frac{1}\omega \bigl[\zeta_c\hat{g}_y+
\zeta_s\hat{g}_z\bigr]\!\right),
\end{equation}
where we have introduced the coefficients
\begin{equation}
\label{osccoeff} \zeta_c=\sqrt{2}\frac{\lambda Q_0(a+a^\dag)}
{\sinh\lambda Q_0}\cos\!\left(\!\omega+\frac{3 Q_0}{4\omega}
\!\right)\!\tau,\quad \zeta_s=\sqrt{2}\frac{\lambda Q_0(a+a^\dag)}
{\sinh\lambda Q_0}\sin\!\left(\!\omega+\frac{3 Q_0}{4\omega}
\!\right)\!\tau,
\end{equation}
which are restricted operators in the oscillator subspace:
$\zeta_c, \zeta_s\sim 1$. It is now seen that $\eta\epsilon$ terms
in Eq. (\ref{hatV}) perform a rotation around $\hat{g}_xa^\dag a$
by a negligibly small angle. We disregard these terms and proceed
with expressing $\exp(-\eta a^\dag a)\hat{A}(\omega)\exp(-\eta
a^\dag a)$ in terms of a single exponent. The fact that we have to
keep terms linear in small parameters only leads to the huge
simplification, as anticipated operators in the single exponent
will not contain interference terms $\propto\eta\cdot(1/\omega)$,
from different small operators. More precisely, from the
perspective of the Campbell-Baker-Hausdorff formula and within the
adopted accuracy, we have
\begin{equation}
\label{tricksingexp}\exp(-\eta a^\dag a)\hat{A}(\omega)\exp(-\eta
a^\dag a)=\exp\!\left(\!\log\left[\exp(-\eta a^\dag
a)\exp(2\lambda Q_0\hat{g}_x^2) \exp(-\eta a^\dag a)\right] +
\frac{1}\omega \bigl[\zeta_c\hat{g}_y+ \zeta_s\hat{g}_z\bigr]\!
\right).
\end{equation}
One however has to take the precaution that in the exact
combination, $\log\left[\exp(-\eta a^\dag
a)\hat{A}(\omega)\exp(-\eta a^\dag a)\right]$, higher terms
$\propto\eta^m(1/\omega)^n$ can arise with large coefficients.
This in turn would mean that the small $\psi$ expansion fails,
i.e., the limit $\psi\to 0$ is singular. Based on the analytic
forms of $Q_0$, $\zeta_c$, and $\zeta_s$, we do not expect such a
failure of the small- $\tau$ expansion, and proceed with Eq.
(\ref{tricksingexp}). We evaluate the argument of the logarithm in
(\ref{tricksingexp}) using the symmetry Eq. (\ref{su2}), as in the
pseudospin sector the three exponents are commutative, and find:
\begin{equation}
\label{hatVres} \hat{V}_{\frac12}(\omega)= \exp(-\epsilon
Q_1\hat{g}_x) \exp\!\!\left(\!2\lambda\bigl[1-8\eta\lambda/9
\bigr] Q_0\hat{g}_x^2 -2\eta a^\dag a +\frac1\omega
\bigl[\zeta_c\hat{g}_y+ \zeta_s\hat{g}_z\bigr]\!\right)
\exp(-\epsilon Q_1\hat{g}_x).
\end{equation}
Taking into account that $\epsilon$ and $\zeta_s$ are odd with
respect to the sign of $\omega$, while $\lambda$, $\zeta_c$, and
$\eta$ are even, we also have
\begin{equation}
\label{hatVnegres} \hat{V}_{\frac12}(-\omega)= \exp(\epsilon
Q_1\hat{g}_x) \exp\!\!\left(\!2\lambda\bigl[1-8\eta\lambda/9
\bigr] Q_0\hat{g}_x^2 -2\eta a^\dag a -\frac1\omega
\bigl[\zeta_c\hat{g}_y- \zeta_s\hat{g}_z\bigr]\!\right)
\exp(\epsilon Q_1\hat{g}_x).
\end{equation}
Within the given approximation, $\eta\lambda=\epsilon^2/4$ is
small and should be discarded. To find the whole time evolution
operator
$\bar{U}=\hat{V}_{\frac12}(\omega)\hat{V}_{\frac12}(-\omega)$, Eq.
(\ref{UbarOnly}), we multiply Eqs. (\ref{hatVres}) and
(\ref{hatVnegres}):
\begin{equation}
\label{Ubarsmall} \bar{U}= \exp(-\epsilon Q_1\hat{g}_x)
\exp\!\!\left(\!2\lambda Q_0\hat{g}_x^2 -2\eta a^\dag a
+\frac1\omega \bigl[\zeta_c\hat{g}_y+
\zeta_s\hat{g}_z\bigr]\!\right) \exp\!\!\left(\!2\lambda
Q_0\hat{g}_x^2 -2\eta a^\dag a -\frac1\omega
\bigl[\zeta_c\hat{g}_y-
\zeta_s\hat{g}_z\bigr]\!\right)\exp(\epsilon Q_1\hat{g}_x).
\end{equation}
We now need to express the product of the two exponents in the
middle of Eq. (\ref{Ubarsmall}) in terms of a single exponent. In
this single exponent, the largest will apparently be the term,
$4\lambda Q_0\hat{g}_x^2$. Except for this, we have to find terms
linear in $\eta$, $\zeta_c/\omega$, and $\zeta_s/\omega$, and
ignore the ones that contain products or higher powers of these
small quantities. Particularly, if we denote by $\hat{Q}_c$ the
operator of the resulting single exponent linear in
$\zeta_c/\omega$, from the Campbell-Baker-Hausdorff formula we
find that the very same operator emerges in a simpler situation,
as
\begin{equation}
\label{simpsitl} \exp\!\!\left(\!2\lambda Q_0\hat{g}_x^2
+\frac1\omega\zeta_c\hat{g}_y\!\right)\exp \left(\!2\lambda
Q_0\hat{g}_x^2 -\frac1\omega\zeta_c\hat{g}_y\!\right)
=\exp\left(\!4\lambda Q_0\hat{g}_x^2+ \hat{Q}_c+ {\cal O}\bigl(
1/\omega^2\bigr) \!\right)
\end{equation}
We evaluate this product utilizing the algebra of Pauli matrices
as above, and find:
\begin{equation}
\label{Qyc}  \hat{Q}_c=2\frac{\zeta_c}\omega \{\hat{g}_x\hat{g}_z+
\hat{g}_z\hat{g}_x\}\tanh \lambda Q_0.
\end{equation}
Evaluating the remaining small, $\propto\eta$ and $\zeta_s/\omega$
operators, and applying the rotation with $\exp(\epsilon
Q_1\hat{g}_x)$, we get
\begin{equation}
\label{Ubarfin} \bar{U}= \exp\!\!\left(\!4\lambda Q_0\hat{g}_x^2
-4\eta a^\dag a -4\epsilon\lambda Q_0\hat{g}_x +\frac2\omega
\!\bigl[\zeta_c\tanh\lambda Q_0\{\hat{g}_x\hat{g}_z+
\hat{g}_z\hat{g}_x\}
 +\zeta_s\hat{g}_z\bigr]\!\right).
\end{equation}
Let's now consider the exponent in Eq. (\ref{Ubarfin}) as the
operator $4\lambda Q_0\hat{g}_x^2$, perturbed with three smaller
operators, $\sim (\epsilon\lambda), \eta$, and $1/\omega$. This
unperturbed operator is diagonal in the pseudospin sector, while
the last term in the square brackets does not have a diagonal in
that sector. Therefore the latter term, representing correction
due to $W_0$, contributes $\sim1/\omega^2$ in the eigenvalue of
$\bar{U}$ (contributions as small as $\eta/\omega$ or
$\epsilon\lambda/\omega$ are also excluded). This statement holds
down to $\lambda\gtrsim \eta$ or $1/\omega$. When $\lambda$
crosses over to smaller values, Eq. (\ref{Ubarfin}) is formally no
longer valid. For small $\lambda$, however, $\bar{U}$ can be
simply found by adding up the small exponents in Eq. (\ref{hatV}),
as commutators amongst them are now suppressed at least
quadratically. The result
\begin{equation}
\label{singexp2} \bar{U}= \exp\!\!\left(\!4\lambda Q_0\hat{g}_x^2
-4\eta a^\dag a +
\frac{2\sqrt{2}}{\omega}\hat{g}_z(a+a^\dag)\sin\!\left(\!\omega+\frac{3
Q_0}{4\omega} \!\right)\!\tau \right),
\end{equation}
nonetheless, coincides with the small-$\lambda$ limit of
(\ref{Ubarfin}). Here the last term corresponds to $W_0$. While
$e_3$ is the eigenvalue of Eq. (\ref{singexp2}) in the pseudospin
space, owing to $\hat{g}_ze_3=0$ the last term does not contribute
in the matrix element $\langle0_z|\bar{U}^N|0_z\rangle$ at all
(recall that $|0_z\rangle\propto e_3$). Meanwhile, the remaining
two terms in Eq. (\ref{singexp2}) represent $U_0$, so that in this
limit the result Eq. (\ref{resulttext}) becomes even more
accurate.

\section{}

In this Appendix we calculate the $\sim1/\omega^2$ correction to
the matrix element, Eq. (\ref{SWres}). Our starting point is Eq.
(\ref{viaUbar}), which contains all secular terms
$\sim1/\omega^2$. We extend the expansion of $\bar{U}$ to the
second order and write:
\begin{equation}
\label{secorU}\bar{U}= U_0+ \frac{1}\omega\, U_1+ \frac1{\omega^2}
\,U_2,
\end{equation}
where $U_1$ is given by Eq. (\ref{perturb}), which we rewrite here
as
\begin{eqnarray}
\label{U1re} U_1\!\!\!\!\!&&= \sqrt{2}\, e^{-\tau
\hat{h}_{+}}\!\left \{ \left[(a+ a^\dag)
\hat{g}_y\right]\!e^{-2\tau \hat{h}_{-}}
-e^{-2\tau \hat{h}_{-}}\!\left[(a+
a^\dag)\hat{g}_y\right] \right \}e^{-\tau \hat{h}_{+}},
\end{eqnarray}
while $U_2$ is found from Eq. (\ref{UbarOnly}) to be
\begin{eqnarray}
\label{U2} U_2\!\!\!\!\!&&= e^{-\tau \hat{h}_{+}}\!\left \{
\left[(a+ a^\dag) \hat{g}_y\right]^2\!e^{-2\tau \hat{h}_{-}} +
e^{-2\tau \hat{h}_{-}}\!\left[(a+ a^\dag)\hat{g}_y\right]^2
-2\,\left[(a+ a^\dag) \hat{g}_y\right]e^{-2\tau
\hat{h}_{-}}\left[(a+ a^\dag) \hat{g}_y\right]\right \}e^{-\tau
\hat{h}_{+}}.
\end{eqnarray}
For the matrix element Eq. (\ref{viaUbar}) we have:
\begin{equation}
\label{newME} \langle0_z|\bar{U}^N |0_z\rangle = \langle0_z|U_0^N
|0_z\rangle+\frac1{\omega^2} \langle0_z|L_N |0_z\rangle.
\end{equation}
Here we introduced the operator,
\begin{equation}
\label{LN} L_N=\sum_{ m_i} U_0^{m_1} U_1 U_0^{m_2} U_1 U_0^{m_3}
+\sum_{n_i} U_0^{n_1}U_2U_0^{n_2},
\end{equation}
where the sums imply $m_1+m_2+m_3=N-2$, and $n_1+n_2=N-1$,
respectively. We first look at the limit of static bath, $\rho\to
0$. It is easy to see that in this limit $\langle0_z|U_0^N
|0_z\rangle=1$. Besides, in this limit $\langle0_z|\bar{U}^N
|0_z\rangle$ can also be found; a simple calculation yields the
form, $\langle0_z|\bar{U}^N |0_z\rangle=1-\kappa/\omega^2$, with a
positive $\kappa<1$, which depends on $N$ only weakly. Therefore,
for the matrix element of $L_N$ we have
\begin{equation}
\label{LNsmrt} \langle0_z|L_N |0_z\rangle=(\rho\tau)l_N-\kappa,
\end{equation}
where $l_N$ incorporates all the secular terms, accumulating with
$N$. This simple analysis suggests that the correction is
suppressed for small $\rho\tau$. We thus restrict our
consideration to $\rho\tau\lesssim 1$. As we keep the short-$\tau$
regime, the latter condition also means a fast bath,
$\rho\gg1/\omega$.

To find the matrix element $\langle0_z|L_N |0_z\rangle$, we turn
to the eigenstates of $U_0$ in the oscillator sector, i.e., the
eigenstates of $V_{\pm}$, denoted by $|\chi^{\pm}_n\rangle$ and
explicitly presented in Eq. (\ref{eigfunc}). We notice that in
this domain of parameters $|\chi^{\pm}_n\rangle$ are very close to
the oscillator states $|n\rangle$. It follows directly from Eq.
(\ref{eigfunc}), where we have $q_y=0$, and $\nu$ close to unity,
$\nu\approx1+\tau^2/6\omega^2+i\rho\tau^2/\omega$.
Particularly, this fact leads to the relation,
$\langle0_z|U_0^N|0_z\rangle\approx[\langle0_z|U_0|0_z\rangle]^N$,
which entails the exponential dependence Eq. (\ref{smallesttau}).
Utilizing this fact and Eqs. (\ref{03viaXpm}), (\ref{U0viaV}) we
write:
\begin{eqnarray}
\label{summand1} &&\langle0_z|U_0^{m_1} U_1 U_0^{m_2} U_1
U_0^{m_3} |0_z\rangle \simeq\frac12\sum_{\sigma,\mu=\pm}\bigl[
\langle0|V_{\sigma} |0\rangle\bigr]^{m_1}\langle
X_{\sigma}|\langle0| U_1 U_0^{m_2} U_1
|0\rangle|X_{\mu}\rangle\bigl[\langle0|V_{\mu}
|0\rangle\bigr]^{m_3},\\
&&\langle0_z|U_0^{n_1} U_2 U_0^{n_2} |0_z\rangle
\simeq\frac12\sum_{\sigma,\mu=\pm}\bigl[ \langle0|V_{\sigma}
|0\rangle\bigr]^{n_1}\langle X_{\sigma}|\langle0|U_2|0\rangle
|X_{\mu}\rangle\bigl[\langle0|V_{\mu} |0\rangle\bigr]^{n_2}.
\end{eqnarray}
Similarly, in the domain of parameters under consideration, the
eigenvectors of $\hat{h}_{\pm}$ in the oscillator sector nearly
coincide with the oscillator eigenstates. We evaluate the
remaining matrix elements of $U_1 U_0^{m_2} U_1$ and $U_2$ from
Eqs. (\ref{U1re}) and (\ref{U2}), where the exact eigenvalues of
$\hat{h}_{\pm}$ are kept, while the eigenvectors are replaced with
the corresponding oscillator states. The resulting matrix element
of $L_N$ reads:
\begin{equation}
\label{Correction0}\langle0_z|L_N |0_z\rangle\simeq
-2e^{-N\vartheta} \cos^2\bigl(\omega+1/4\omega\bigr)\tau
\tanh(\tau\rho-\vartheta)
\left\{N+\frac{1-e^{-4N(\rho\tau-\vartheta)}}{2\sinh2(\tau\rho-\vartheta)}\right\},
\end{equation}
where we have introduced  small $\vartheta\ll1$ through
$e^{-\vartheta}= \langle0_z|U_0|0_z\rangle$. With
$\rho\tau\lesssim1$, the last term of Eq. (\ref{Correction0}) can
be neglected. Approximations made in course of this calculation,
consisting in the substitution of eigenvectors of $U_0$ and
$\hat{h}_{\pm}$ by corresponding oscillator states, lead to the
error, $\sim (\tau/\omega)^2$ or $1/(\rho\omega)^2$. As
$\vartheta$ has the same order of magnitude, we discard it from
the argument of hyperbolic function.
The correction to $\langle S^z(N)\rangle$ thus acquires the
compact form,
\begin{equation}
\label{Correction} \Delta\langle S^z(N)\rangle=
-\frac{2N}{\omega^2}\langle0_z|U_0^N |0_z\rangle \tanh(\tau\rho)
\cos^2\bigl(\omega+1/4\omega\bigr)\tau .
\end{equation}

\section{}

In this Appendix, we evaluate the quantum mechanical average
$\langle 0_z|U_\lambda^N|0_z\rangle$ for $U_\lambda=e^{-\tau
\hat{h}_{\lambda +}}e^{-2\tau \hat{h}_{\lambda -}} e^{-\tau
\hat{h}_{\lambda +}}$, with $\hat{h}_{\lambda\pm}$ given by
Eq.~(\ref{hsmHyp}), and find $\langle S^z_\lambda(N)\rangle$ for
$\lambda\ll\omega$.

In full analogy with Eq. (\ref{me1}) for $U_0$, we have $\langle
0_z|U_\lambda^N|0_z\rangle=\text{Re}\langle0| V_\lambda^N
|0\rangle$, where $V_\lambda=e^{-\tau h_{\lambda +}}e^{-2\tau
h_{\lambda -}} e^{-\tau h_{\lambda +}}$, and
\begin{equation}
\label{hHyp} h_{\lambda\pm}= \rho\, a^\dag a \mp i\omega \mp
i\frac{(a+ a^\dag+\lambda)^2}{4\omega}.
\end{equation}
We relate $V_\lambda$ to the operator $V(\omega)=e^{-\tau
h_{0+}}e^{-2\tau h_{0-}} e^{-\tau h_{0 +}}$ with $h_{0
\pm}=h_{\lambda \pm}|_{\lambda=0}$, which was already introduced
in Appendix \ref{AppB}, cf. Eq. (\ref{Vpm}). This is done by
exploiting the relation,
\begin{equation}
\label{Bt} e^{-\tau h_{\lambda+}} =e^{-\tau\lambda\rho\phi/2}
e^{-\phi(a^\dag-a)}e^{-\tau h_{0+}}e^{\phi(a^\dag-a)}, \qquad
\phi=\frac{\lambda}{2(1+ i\rho\omega)},
\end{equation}
and its counterpart for $h_{\lambda-}$, where $\phi$ is replaced
with its complex conjugate, $\phi^*$. We get:
\begin{equation}
\label{Vl} V_\lambda= e^{-\tau\lambda\rho(\phi+\phi^*)}
e^{-\phi(a^\dag-a)}\underbrace{e^{-\tau h_{0+}}
e^{\eta(a^\dag-a)}e^{-2\tau h_{0-}}e^{-\eta(a^\dag-a)} e^{-\tau
h_{0+}}}_{V^\prime} e^{\phi(a^\dag-a)},\qquad \eta=\phi-\phi^*.
\end{equation}
From the commutation relations between $h_{0+}$, $a^\dag$, and
$a$, one can infer the identity,
\begin{equation}
\label{rotrot} e^{-\tau h_{0+}}(a^\dag-a)e^{\tau h_{0+}}=
(a^\dag-a)\cosh\tau P-(a^\dag+a)\frac P\rho\sinh\tau P,
\end{equation}
where $P=\sqrt{\rho^2-i\rho/\omega}$ was also introduced in
Appendix \ref{AppB}. Using the exponential form of $V(\omega)$,
Eq. (\ref{fullcycle}), we rewrite $V^\prime$ as
\begin{equation}
\label{Vprime} V^\prime=e^{\eta\left[ (a^\dag-a)\cosh\tau
P-(a^\dag+a)(P/\rho)\sinh\tau P\right]}\, e^{-\psi(\vec{q}\vec{K})
+2\tau\rho}\, e^{-\eta\left[ (a^\dag-a)\cosh\tau
P+(a^\dag+a)(P/\rho)\sinh\tau P\right]}.
\end{equation}
Next we notice the commutation relations, $[(\vec{q}\vec{K}),
C_\pm]=\pm C_\pm$, where $C_\pm=(a^\dag+a)\pm (q_z-iq_x)
(a^\dag-a)$. At the same time, $[C_-, C_+]=4(q_z-iq_x)$ is a
scalar, so that the first and third exponents in Eq.
(\ref{Vprime}) are easily expressed via $C_\pm$:
\begin{equation}
\label{Vprimexy} V^\prime=e^{x C_-}e^{y C_+}\,
e^{-\psi(\vec{q}\vec{K}) +2\tau\rho}\, e^{x C_+}e^{y C_-},
\end{equation}
where
\begin{equation}
\label{xy} x=-\frac\eta2\bigl[(q_z+iq_x)\cosh\tau P+
(P/\rho)\sinh\tau P\bigr],\qquad y=\frac\eta2\bigl[(
q_z+iq_x)\cosh\tau P- (P/\rho)\sinh\tau P \bigr].
\end{equation}
It is now straightforward to cast $V^\prime$ into the form of a
single exponent from Eq. (\ref{Vprimexy}), using the identity,
$\exp X \exp Y=\exp\{X+sY/(1-e^{-s})\}$, which follows for any two
operators $X$ and $Y$ with the commutator, $[X,Y]=sY$. Introducing
\begin{equation}
\label{eps} \epsilon=\eta\bigl[\cosh\tau P-(q_z-iq_x)
(P/\rho)\sinh\tau P\coth(\psi/2)\bigr],
\end{equation}
the resulting entangled form of $V^\prime$ reads:
\begin{equation}
\label{Vent} V^\prime=\exp\left(-\psi\Bigl[(\vec{q}\vec{K})
-(q_z+iq_x)\epsilon\,(a^\dag+a)+(q_z+iq_x)\epsilon^2\Bigr]
+2\tau\rho +2(x+y)\epsilon \right).
\end{equation}
The combination in rectangular brackets of the exponent Eq.
(\ref{Vent}) is equal to
$\exp[\epsilon\,(a^\dag-a)](\vec{q}\vec{K})
\exp[-\epsilon\,(a^\dag-a)]$, so that using Eq. (\ref{Vl}) we get:
\begin{equation}
\label{Vlfin} V_\lambda= e^{2(x+y)\epsilon
-\tau\lambda\rho(\phi+\phi^*)}
e^{-(\phi-\epsilon)(a^\dag-a)}\,V(\omega)\,
e^{(\phi-\epsilon)(a^\dag-a)}.
\end{equation}
From this relation, the sought expectation value, $\langle0|
V_\lambda^N |0\rangle$, is expressed via the one involving
$V(\omega)$ as
\begin{equation}
\label{VlV} \langle0| V_\lambda^N |0\rangle = e^{N[2(x+y)\epsilon
-\tau\lambda\rho(\phi+\phi^*)]}
\langle0|e^{-(\phi-\epsilon)(a^\dag-a)}\,V^N(\omega)\,
e^{(\phi-\epsilon)(a^\dag-a)}|0\rangle.
\end{equation}
The action of translation $e^{(\phi-\epsilon)(a^\dag-a)}$ on a
wave function results in a shift of its argument by
$-\sqrt{2}(\phi-\epsilon)$. Consequently, the normalized vector
$e^{(\phi-\epsilon)(a^\dag-a)}|0\rangle$ is expanded in terms of
the eigenvectors of $V(\omega)$, Eq. (\ref{eigfunc}), by
\begin{equation}
\label{expanded} e^{(\phi-\epsilon)(a^\dag-a)}|0\rangle=
\sum_{n\geq0} F_n|\chi_n\rangle,\qquad
F_n=\frac{\nu^{\frac14}}{\sqrt{\pi2^nn!}}\int d\xi
e^{-\frac12\bigl([\xi-\sqrt{2}(\phi-\epsilon)]^2
+\nu\xi^2\bigr)}H_n(\sqrt{\nu}\xi),
\end{equation}
where $H_n$ are Hermite polynomials. Carrying out the integration
yields:
\begin{equation}
\label{Fn}F_n=\frac{\nu^{\frac14}}{\sqrt{(1+\nu)2^{n-1}n!}}
\left(\frac{1-\nu}{1+\nu}\right)^{\frac n2}\exp\!\left(-\frac{
\nu[\phi-\epsilon]^2}{1+\nu} \right)H_n\!\left(
\sqrt{\frac{2\nu}{1-\nu^2}}[\phi-\epsilon]\right).
\end{equation}
This relation extends Eq. (\ref{cn}) to non-zero values of
$\lambda$. The matrix element Eq. (\ref{VlV}) is expressed in
terms of the sum, $\sum_{n\geq 0}F_n^2e^{-N\psi n}$. The analytic
expression for the sum is possible by virtue of the relation,
$\sum_{n\geq
0}H_n^2(x)u^n/n!=(1-4u^2)^{-1/2}\exp\bigl[4ux^2/(1+2u)\bigr]$. We
find:
\begin{equation}
\label{Ulans} \langle0| U_\lambda^N |0\rangle = \text{Re}\!
\left[\! \langle0| V(\omega)^N |0\rangle \exp\!
\left(\![2(x+y)\epsilon -\tau\lambda\rho(\phi+\phi^*)]N
-(\phi-\epsilon)^2\frac{2\sinh \frac N2\psi}{\sinh \frac N2 \psi
+(q_z-iq_x)\cosh \frac N2\psi} \right)\right],
\end{equation}
where $\langle0| V(\omega)^N |0\rangle$ is factored out by using
Eq. (\ref{me2}). To visualize this behavior, we make use of
Eqs.~(\ref{fcm1})-(\ref{fcm}) and derive the short-$\tau$
asymptotes,
\begin{equation}
\label{xyeasymp} 2(x+y)\epsilon
-\tau\lambda\rho(\phi+\phi^*)\simeq
-\frac{\lambda^2\rho\tau^3}{3\omega^2},\qquad \phi-\epsilon\simeq
\frac{\lambda\tau^2}{4\omega^2}\bigl(2/3-i\rho\omega\bigr).
\end{equation}
We also notice that for typical values of parameters the last term
in the exponent of Eq. (\ref{Ulans}) saturates with $N$ rapidly.
Effectively, for relevant values of $N$ one has:
\begin{equation}
\label{combasymp} \frac{\sinh \frac N2\psi}{\sinh \frac N2 \psi
+(q_z-iq_x)\cosh \frac N2\psi} \simeq \frac1{1+q_z-iq_x}\approx
\frac12.
\end{equation}
Thus the last term  of Eq. (\ref{Ulans}) is $\propto\tau^4$ and
does not accumulate with $N$. Neglecting this term, and using Eqs.
(\ref{smallesttau}), (\ref{Sl}), we arrive at the short-$\tau$
form Eq. (\ref{withHyp}).

\end{widetext}


\begin{references}
\bibitem{footnote} A detailed analysis shows that this
contribution is $\sim 1/\omega^2$.
\end{references}
\end{document}